\documentclass[11pt,a4paper]{article} 
\pdfoutput=1
\usepackage{jheppub}                  

\usepackage{graphicx}
\usepackage{dsfont}

\usepackage[normalem]{ulem}	
\newcommand{\dMsq}{\Delta m^2_{ \text{ren} }}
\newcommand{\am}{a}

\def\lsim{\raise0.3ex\hbox{$\;<$\kern-0.75em\raise-1.1ex
\hbox{$\sim\;$}}}
\def\gsim{\raise0.3ex\hbox{$\;>$\kern-0.75em\raise-1.1ex
\hbox{$\sim\;$}}}

\title{ Simple and Compact Expressions for Neutrino Oscillation Probabilities in Matter \\ }
\author{Hisakazu Minakata$^{1}$}
\author{Stephen~J.~Parke$^{2}$}
\affiliation{
$^1$Instituto de F\'{\i}sica, Universidade de S\~ao Paulo, C.\ P.\
66.318, 05315-970 S\~ao Paulo, Brazil \\
$^2$Theoretical Physics Department, Fermi National Accelerator Laboratory, P.\ O.\ Box 500, Batavia, IL 60510, USA \\ 
}

\abstract{ 
We reformulate perturbation theory for neutrino oscillations in matter with an expansion parameter related to the ratio of the solar to the atmospheric $\Delta m^2$ scales. Unlike previous works, we use a renormalized basis in which certain first-order effects are taken into account in the zeroth-order Hamiltonian. 
We show that the new framework has an exceptional feature that leads to the neutrino oscillation probability in matter with the same structure as in vacuum to first order in the expansion parameter. It facilitates immediate physical interpretation of the formulas, and makes the expressions for the neutrino oscillation probabilities extremely simple and compact. We find, for example, that the $\nu_e$ disappearance probability at this order is of a simple two-flavor form with an appropriately identified mixing angle and $\Delta m^2$. 
More generally, all the oscillation probabilities can be written in the universal form with the channel-discrimination coefficient of $0,~\pm1$ or simple functions of $\theta_{23}$. Despite their simple forms they include all order effects of $\theta_{13}$ and all order effects of the matter potential, to first order in our expansion parameter. 

} 


\emailAdd{minakata@fmail.if.usp.br}
\emailAdd{parke@fnal.gov}

\begin{document} 

\maketitle

\section{Introduction}
\label{sec:introduction}

Neutrino oscillation based on the standard three-flavor scheme provides the best possible theoretical framework available to date to describe most of the experimental results obtained in the atmospheric, solar, reactor, and the accelerator neutrino experiments. Although numerically calculated neutrino oscillation probabilities suffice to analyze experimental data, understanding of the framework, in particular the one in matter \cite{Wolfenstein:1977ue}, has not yet reached  a sufficient level, in our opinion. Under the assumption of uniform matter density distribution, the exact expressions of the eigenvalues, mixing angles, and the oscillation probabilities in matter have been obtained \cite{Barger:1980tf,Zaglauer:1988gz,Kimura:2002wd}.  Yet, the results for these quantities are generally too complicated to facilitate understanding of the structure of the three flavor neutrino oscillations in matter primarily due to the complexities of solving the cubic eigenvalue characteristic equation. For a recent comprehensive treatment of neutrino oscillation in the matter, see ref.~\cite{Blennow:2013rca}.

Analytic approaches to the neutrino oscillation phenomenon, so far, are mostly based on variety of perturbative frameworks. If the matter effect is small one can treat it as a small perturbation \cite{Arafune:1997hd}. In the environments in which the matter effect is comparable to the vacuum mixing effect, the only available small expansion parameter known to us is the ratio of the solar-scale $\Delta m^2_\odot$ to the atmospheric-scale $\Delta m^2_\oplus$, $  \Delta m^2_\odot / \Delta m^2_\oplus \simeq 0.03$. $\sin \theta_{13}$ has been often used as an expansion parameter (there are enormous number of references, see e.g., \cite{Cervera:2000kp}), but it is now known that its value is not so small, $\sin \theta_{13} \simeq 0.15$, which is of the order of $\sqrt{ \Delta m^2_\odot / \Delta m^2_\oplus }$. Moreover, expansion around $\sin \theta_{13}=0$ misses the physics of the resonance 
which exists at an energy around $E \sim 10$ GeV for earth densities. Therefore, it appears that the suitable perturbative framework is the one with the unique expansion parameter  $  \Delta m^2_\odot / \Delta m^2_\oplus$. This framework was indeed examined in the past, to our knowledge in refs.~\cite{Arafune:1996bt,Cervera:2000kp,Freund:2001pn,Akhmedov:2004ny}. 

In this paper, we present a new framework of perturbative treatment of neutrino oscillation in matter. We follow the reasoning stated above which led to identification of the unique expansion parameter $\epsilon \approx \Delta m^2_\odot / \Delta m^2_\oplus$. But, unlike the preceding works, we use a ``renormalized basis'' as the basis of perturbation. That is, we absorb certain terms of order $\epsilon$ to our ``zeroth-order'' Hamiltonian around which we perturb. Or, in other word, we take the zeroth-order eigenvalues in matter such that it matches the exact eigenvalues to order $\epsilon$, see section~\ref{sec:more-about}. 
We will show that use of the renormalized basis makes the structure of the perturbation theory exceptionally transparent, as we will explain in the next section. It allows us to obtain simple, elegant and compact expressions for the oscillation probabilities, which have a universal form even to first order in our expansion parameter. 

For example, $\nu_{e}$ survival probability takes the form to order $\epsilon$ as
\begin{eqnarray}
P(\nu_e \rightarrow \nu_e)  &=&
1 -  \sin^2 2\phi  ~\sin^2 \frac{ (\lambda_{+} - \lambda_{-} ) L }{4E} 
\label{eq:P-ee-demo}
\end{eqnarray}
where $\phi$ is $\theta_{13}$ in matter, and $\lambda_{\pm}$ denote the eigenvalues of the states which participate the 1-3 level crossing. Despite its extremely simple form, $P(\nu_e \rightarrow \nu_e)$ in (\ref{eq:P-ee-demo}) takes into account all order effects of $\theta_{13}$ and  the matter potential. Since we will only consider terms up to order $\epsilon$, in this paper,  our results here are not applicable to the region near the solar  MSW resonance \cite{Mikheev:1986gs,Wolfenstein:1977ue}. Our perturbative framework will be called as the ``renormalized helio-perturbation theory'' in the rest of this paper. 

The section plan of this paper is somewhat unusual: in the next section~\ref{sec:structure} we summarize the key features of the perturbative framework we develop in this paper. Then, in section~\ref{sec:results} we describe the principle results of this work including the oscillation probabilities for all channels in matter. This section does not describe the derivations but provides a self contained summary of the results of this paper. Following this section, see section~\ref{sec:formulation}, we present a systematic exposition of our perturbative framework and how the results of the section~\ref{sec:results} are derived. In the appendices~\ref{sec:calculating-S-matrix}, \ref{sec:oscillation-P}, and~\ref{sec:relations} we present, respectively, calculational details of the $S$ matrix, the results of oscillation probabilities to order $\epsilon$ in the standardized form, and useful relationships to verify the equivalence of various expressions. 

\section{Structure of the neutrino oscillation probabilities in vacuum and in uniform matter} 
\label{sec:structure}

We start by giving a precise definition of what we mean by ``simplicity and compactness'' of the expressions for the neutrino oscillation probabilities. 
Then, we explain the need for reformulating perturbative treatment, followed by outlining ``what is new'' in this paper.

\subsection{Simplicity and Compactness}

In vacuum and in matter with constant density, it is well known that the neutrino oscillation probabilities for $\nu_\beta \rightarrow \nu_\alpha$ for {\it three-flavor} mixing ($i, j = 1, 2, 3$) can be written in the following form\footnote{
We have used for the last term in (\ref{eq:P-ba-general})
${\mbox Im} [V_{\alpha 1} V_{\beta 1}^* V_{\alpha 2}^* V_{\beta 2}] = {\mbox Im} [V_{\alpha 2} V_{\beta 2}^* V_{\alpha 3}^* V_{\beta 3}]  = - {\mbox Im} [V_{\alpha 1} V_{\beta 1}^* V_{\alpha 3}^* V_{\beta 3}] $ which follows from unitarity, and the identity (\ref{eq:sss-identity}). 
} 
\begin{eqnarray}
P(\nu_\beta \rightarrow \nu_\alpha) &=& 
\left| ~\sum_{i=1}^{3} V_{\alpha i} V^*_{\beta i}  ~e^{-i\frac{\lambda_iL}{2E}} ~\right|^2 = 
\delta_{\alpha \beta}  - 4 \sum_{j > i}  {\mbox Re}[V_{\alpha i} V_{\beta i}^* V_{\alpha j}^* V^{ }_{\beta j}]
\sin^2 \frac{ (\lambda_{j} - \lambda_{ i}) L}{ 4E } 
\nonumber\\
&&  \hspace{-1cm}+~
8 ~{\mbox Im} [V_{\alpha 1} V_{\beta 1}^* V_{\alpha 2}^* V^{ }_{\beta 2}]  
~\sin \frac{ (\lambda_{3} - \lambda_{2}) L}{ 4E } 
\sin \frac{ (\lambda_{2} - \lambda_{1}) L}{ 4E } 
\sin \frac{ (\lambda_{1} - \lambda_{3}) L}{ 4E }.
\label{eq:P-ba-general}
\end{eqnarray}
The flavor mixing matrix elements, $V_{\alpha i}$, relate the flavor states, $\nu_{\alpha}$, to the eigenstates of the Hamiltonian, $\nu_{i}$, with eigenvalues $\lambda_{i} / 2E$ by $\nu_{\alpha} = V_{\alpha i} \nu_{i}$. Equation (\ref{eq:P-ba-general}) applies both in vacuum and in uniform matter since the matrix $V$ diagonalizes the full Hamiltonian, which includes the Wolfenstein matter potential, $a$ \cite{Wolfenstein:1977ue}. In vacuum, $\lambda_{i}=m^2_{i}$ where $m_{i}$ denotes the mass of the $i$-th neutrino state and $V= U_{\rm MNS}$. 

Notice that the identical form of the oscillation probabilities in vacuum and in matter, eq.~(\ref{eq:P-ba-general}), apart from replacement of $\Delta m^2_{ij}$ by $(\lambda_i -\lambda_j)$, implies that physical interpretation of the formulas in matter is as transparent as in vacuum.  

What is less well appreciated is that the expressions of the oscillation probabilities in (\ref{eq:P-ba-general}) are maximally simple and compact. That is, they contain 5 (including the constant term) functions of $L/E$ which are linearly independent. This property can be easily verified by showing the Wronskian is nonvanishing, which implies that none of the functions can be written in terms of the other functions for all $L/E$. 
Hence, in a true mathematical sense, eq.~(\ref{eq:P-ba-general}) and it's equivalents give the simplest and most compact form for the complete set of three flavor oscillation probabilities in vacuum and in uniform matter.\footnote{ 
There are equivalent ways to write this set of oscillation probabilities with 5 independent $L/E$ functions, e.g one could use the trigonometric identity $2 \sin^2 x=1-\cos2x$ to replace the $\sin^2 x$.
Any other way of writing these oscillation probabilities will have 5 or more linearly independent functions, e.g., using the identity 
(\ref{eq:sss-identity}) increases the number of independent functions by 2. 
}

A special feature of the oscillation probabilities which is also worth noting is that each term in eq.~(\ref{eq:P-ba-general}) factorizes into the characteristic $\sin \left[ (\lambda_{j} - \lambda_{i}) L / 4E  \right]$ factor and the products of the $V$ matrix elements which control the amplitude of the oscillation. Both the eigenvalues, $\lambda_{i}$ and the matrix elements of $V$ are independent of the baseline, $L$, but are functions of the mixing angles, $\theta$'s, the  $\Delta m^2_{ji}$, and the product of the energy of neutrino times the matter density via the matter potential.\footnote{
$a \propto \rho E$ with $\rho$ being the matter density.
The $V$ matrix elements are determined solely by the elements of the Hamiltonian matrix. It is easy to observe this feature by using the method developed in ref.~\cite{Kimura:2002wd}, which is valid for nonuniform matter density as well. 
 }
CP and T violation is described by the last term in (\ref{eq:P-ba-general}), which has the universal, channel-independent form in the three neutrino mixing.

\subsection{Raison d'~\^etre and the requirements for perturbative treatment}

With the simplest form of the oscillation probability eq. (\ref{eq:P-ba-general}) and by knowing both the exact form of the eigenvalues, $\lambda_i/2E$, and the elements of the $V$ matrix,  see  \cite{Zaglauer:1988gz}, one might expect we have all that  is needed for theoretical discussions. Unfortunately, the analytic expressions for these eigenvalues, $\lambda_i/2E$,  as well as the $V$ matrix elements are notoriously complex and give no analytic insight into the oscillation physics in uniform matter. This is true even when one of $\Delta m^2_{21}$, $\sin \theta_{13}$, $\sin \theta_{12}$ or $a$ is set equal to zero. 
In any one of these limits, the characteristic equation for the eigenvalues factorizes. But, the form of the general solution does not simplify trivially to yield the correct eigenvalues, even though it must.   The structure of the general solutions of the cubic characteristic equation is the root cause of this rather unusual behaviour. 
Hence, there is a need for a reformulated perturbative framework so that  we can obtain approximate but much simpler  expressions for the eigenvalues, $\lambda_{i}$, and the mixing matrix elements, $V_{\alpha i}$, which provide the necessary physics insight.

In this paper, we formulate perturbation theory by which we can calculate the eigenvalues $\lambda_{i}$ and the elements of $V$ matrix as a simple power series of the small expansion parameter, a renormalized $\Delta m^2_\odot / \Delta m^2_\oplus$. At the same time the structure-revealing form of the oscillation probabilities (\ref{eq:P-ba-general}) is kept intact. While the existing frameworks do not satisfy the latter requirement, the key to the success in our case is due to the correct decomposition of the Hamiltonian into the unperturbed and the perturbed terms. Use of the renormalized zeroth-order basis, allows us to correctly determine the eigenvalues to the appropriate order in our expansion parameter. The resultant renormalized eigenvalues and the mixing parameters effectively absorb the additional terms that arise in the conventional perturbative frameworks, as extra functions to the minimally required 5 linearly independent functions. This occurs automatically and we believe that such a framework has never before been formulated.\footnote{
See, however, the clarifying remark at the end of subsection~\ref{sec:comparison}. 
}
Because of the structural simplicity of the $L/E$ dependence, the expressions of the oscillation probabilities calculated by our method are extremely simple and compact, as will be fully demonstrated in the next section.

\section{Results of our perturbative expansion for all oscillation probabilities}
\label{sec:results}

In this section, we describe the main results of this paper without derivations and with only minimal discussion.  In later sections we provide the derivation and more in depth discussions.  We start with the approximate eigenvalues of the Hamiltonian, the approximate neutrino mixing matrix and then give the oscillation probabilities for all channels to first order in the expansion parameter, $\epsilon$, see eq. (\ref{eq:epsilon-def}) for the precise definition.

\subsection{Mass squared eigenvalues  in matter}
\label{sec:lambdas}

In vacuum the three eigenvalues of the full Hamiltonian which governs the neutrino oscillation is given in the form 
$m^2_{i}/2E$, where $m_{i}$ is the mass of $i$-th mass eigenstate of neutrinos,  $i = 1, 2, 3$.  Similarly, in matter we write the three eigenvalues as 
\begin{eqnarray}
\frac{\lambda_{i}}{2E},  \nonumber
\end{eqnarray}
where the state label runs over $i = -, 0,+$ for the approximate Hamiltonian of three flavor mixing system.  
To treat the normal and the inverted mass orderings (NO and IO respectively) in a unified way, we define the eigenvalues as follows\footnote{
We note that the eigenvalues in (\ref{eq:lambda-pm0}) above appear in ref.~\cite{Blennow:2013rca}. See section~\ref{sec:basis} for the derivation and a comment on the treatment in \cite{Blennow:2013rca}.
}
\begin{eqnarray} 
\lambda_{-} &=& 
\frac{ 1 }{ 2 } \left[
\left( \dMsq + a \right) - {\rm sign}(\dMsq) \sqrt{ \left( \dMsq - a \right)^2 + 4 s^2_{13} a \dMsq }
\right] 
+ \epsilon \dMsq s^2_{12},
\nonumber\\[3mm]
\lambda_{0} &=&  c^2_{12} ~\epsilon ~\dMsq,
\label{eq:lambda-pm0}  
\\[3mm]
\lambda_{+} &=& 
\frac{ 1 }{ 2 } \left[
\left( \dMsq + a \right) + {\rm sign}(\dMsq) \sqrt{ \left( \dMsq - a \right)^2 + 4 s^2_{13} a \dMsq }
\right] 
+ \epsilon \dMsq s^2_{12}.
\nonumber 
\end{eqnarray}
In eq.~(\ref{eq:lambda-pm0}), the renomalized $\Delta m^2 \equiv \dMsq$, the expansion parameter $\epsilon$, and the Wolfenstein matter potential \cite{Wolfenstein:1977ue}, $a$, are defined as follows:\footnote{
The following notation is used throughout: $\Delta m^2_{ij} \equiv m^2_{i}-m^2_{j}$, $s_{ij}= \sin \theta_{ij}$ and  $c_{ij}= \cos \theta_{ij}$ where $\theta_{ij}$ are the standard neutrino mixing angles and $G_F$ is the Fermi constant, $N_e$ is the number density of electrons, $E$ is the energy of the neutrino, $Y_e$ the electron fraction and $\rho$ is the density of matter. 
}
\begin{eqnarray} 
\dMsq  & \equiv  & \Delta m^2_{31} - s^2_{12} \Delta m^2_{21}, 
\label{eq:Dm2-ren-def}
\\
  \epsilon  & \equiv & \frac{ \Delta m^2_{21} }{ \dMsq },
\label{eq:epsilon-def}
\\
{\rm and }  \quad \quad  \quad a & = & 2\sqrt{2} G_F N_e E \approx 1.52 \times 10^{-4} \left( \frac{Y_e \rho}{\rm g.cm^{-3}} \right) \left( \frac{E}{\rm GeV} \right) {\rm eV}^2.
\label{matt-potential}
\end{eqnarray}
This choice of $\dMsq$ is crucial to the compact formulas for the oscillation probabilities that will be given in this paper.
Note also that the sign of $\dMsq$ signals the mass ordering, both $\dMsq$ and $\epsilon$ are positive (negative) for NO (IO). However, for both orderings $\epsilon \dMsq = \Delta m^2_{21} > 0$, as required by nature. Notice that $\lambda_{0}$ is the same for the both mass orderings, and when we switch from NO to IO we also switch the  sign in front of the square root in eq.~(\ref{eq:lambda-pm0}).
The nicest feature of the sign choice is that the oscillation probability has a unified expression and the solar resonance is in $\nu_{-}$-$\nu_{0}$ level crossing for the both mass orderings.

$\dMsq$ is equal to the effective atmospheric $\Delta m^2$ measured in a electron (anti-) neutrino disappearance experiment in vacuum,  $\Delta m^2_{ee} \equiv c^2_{12} \Delta m^2_{31}+s^2_{12} \Delta m^2_{32}$ \cite{Nunokawa:2005nx}. This quantity is identical to $\Delta m^2_{ee}$ recently measured by the reactor $\theta_{13}$ experiment \cite{An:2013zwz} up to effects of ${\cal O}(\Delta m^2_{21}/\Delta m^2_{31})^2$. Whether the coincidence between $\Delta m^2_{ee}$ and $\dMsq$ reflects a deep aspect of neutrino oscillation or not will be judged depending upon what happens at second order in $\epsilon$.
This point as well as the relevance of the other effective $\Delta m^2_{\mu \mu}$ \cite{Nunokawa:2005nx}, $\nu_{\mu}$ equivalent of $\Delta m^2_{ee}$, will be discussed in depth in a forthcoming communication. 

\subsection{The mixing angle $\theta_{13}$ and mixing matrix in matter}
\label{sec:phi}

We use the angle $\phi$ to represent the mixing angle $\theta_{13}$ in matter. With the definitions of the eigenvalues (\ref{eq:lambda-pm0}),  the following mass-ordering independent expressions for cosine and sine $2 \phi$ (see section~\ref{sec:hat-basis}) are given by 
\begin{eqnarray} 
\cos 2 \phi &=& 
\frac{ \dMsq \cos 2\theta_{13} - a }{ \lambda_{+} - \lambda_{-} },
\nonumber \\
\sin 2 \phi &=& \frac{ \dMsq \sin 2\theta_{13} }{ \lambda_{+} - \lambda_{-} }.
\label{eq:cos-sin-2phi}
\end{eqnarray}

It is easy to show that $\phi$ goes from $0 \rightarrow ~\pi/2$ as $a$ goes from $-~\infty$ to  $+~\infty$ for the NO and as $a$ goes from $+~\infty$ to  $-~\infty$ for the IO. In vacuum ($a=0$), $\phi= \theta_{13}$ and $\phi=\pi/4$ at the atmospheric resonance, when $a=\dMsq \cos 2 \theta_{13}$, for both mass orderings.

The mixing matrix in matter, $V$, relates the flavor eigenstates, $\nu_e$,  $\nu_\mu$, $\nu_\tau$, to the perturbatively defined matter mass eigenstates, $\nu_-$, $\nu_0$, $\nu_+$ as follows (see section~\ref{sec:V-matrix}): 
\begin{eqnarray}
\left(   \begin{array}{l}
\nu_e \\ \nu_\mu \\ \nu_\tau
\end{array}  \right)
& = &
V~\left(   \begin{array}{l}
\nu_{-} \\ \nu_{ 0}  \\ \nu_{ +} 
\end{array}  \right)   
\label{eq:V-matrix-def}
\end{eqnarray}
where  the matrix $V$ is unitary.  It is convenient to split $V$ into a zeroth order term, $V^{(0)}$, and a first order term, $V^{(1)}$ in our $\epsilon$ expansion, 
\begin{eqnarray}
V\equiv V^{(0)} +  \epsilon V^{(1)},
\label{eq:V}
\end{eqnarray}
where the zeroth order matrix is given by 
\begin{eqnarray}
\hspace*{-0.5cm}
V^{(0)} &=& 
\left[ 
\begin{array}{ccc}
c_\phi  &  0  & s_\phi \\
- s_\phi s_{23} e^{ i \delta} \quad  & c_{23}  & c_\phi s_{23} e^{ i \delta} \\
- s_\phi c_{23} & - s_{23} e^{ - i \delta}  \quad &  c_\phi c_{23} 
\end{array} 
\right]. 
\label{eq:V0}
\end{eqnarray}
with $\delta$ being the CP violating phase, whereas the first order correction is given by
\begin{eqnarray}
\hspace*{-2cm}
 V^{(1)} 
& = & c_{12} s_{12}~ \dMsq
 \left\{ \quad
   \left( \frac{  c_{\left( \phi - \theta_{13} \right)}  }{ \lambda_{-} - \lambda_{0} } \right) 
    \left[ 
~\begin{array}{ccc} 
0& -c_{\phi}  & 0
\\ 
c_{23} 
& s_\phi  s_{23} e^{ i \delta} 
&  0
\\ 
- s_{23} e^{ - i \delta } 
& s_\phi c_{23}  
& 0  \\
\end{array} 
\right] 
\right.
\nonumber \\
  & & \left.  \hspace*{2.5cm}+
\left(\frac{ s_{\left( \phi - \theta_{13} \right)} }{ \lambda_{+} - \lambda_{0} }
\right)
\left[ \begin{array}{ccc} 
0&   -s_{\phi}   & 0
\\ 
0
&-c_\phi s_{23} e^{ i \delta} 
&  c_{23} 
\\ 
0
& - c_\phi c_{23}  
&  -s_{23} e^{ - i \delta }
\\
\end{array} 
\right] 
~~\right\}.
\label{eq:V1-sec2}
\end{eqnarray}
A factorized form of the $V$ matrix is given in eq.~(\ref{eq:Vfactorized}). 

 As an outcome of the consistent perturbative treatment the total $V$ matrix given by (\ref{eq:V}) with (\ref{eq:V0}) and (\ref{eq:V1-sec2}) must be unitary to order $\epsilon$. In fact it is, since the following two conditions are satisfied 
\begin{eqnarray}
V^{(0)} (V^{(0)})^\dagger =1  \quad {\rm and} 
\quad  V^{(0)} (V^{(1)})^\dagger + V^{(1)} (V^{(0)})^\dagger =0.  
\end{eqnarray}
Of course, none of what follows is self consistent without unitarity here.

With the matter eigenvalues, $\lambda$'s , definite by eq.~(\ref{eq:lambda-pm0})  and the matter mixing matrix, $V$, given by eq.~(\ref{eq:V}), simple and compact expressions can be easily derived for the oscillation probabilities in matter for all channels,  to leading order in $\epsilon$, as will be shown in the next section.

\subsection{Compact formulas for the oscillation probabilities in matter}
\label{sec:compact-F}

In this section we start by presenting the shortest path to the oscillation probabilities of the $\nu_{e}$-related channel by using the eigenvalues, $\lambda_{\pm,0}$, and mixing matrix, $V$, given in the previous section to order $\epsilon$. Then, we derive the universal expression of the oscillation probabilities which is applicable to all channels. 

\subsubsection{$\nu_e \rightarrow \nu_e$ disappearance channel}
\label{sec:Pee}

The derivation of the $\nu_{e}$ survival oscillation probability, $P(\nu_e \rightarrow \nu_e)$,  in our renormalized helio-perturbation theory is extremely simple. 
Starting from the general expression 
\begin{eqnarray}
P(\nu_e \rightarrow \nu_e)  &=& 1- 4 |V_{e+}|^2 |V_{e-}|^2 \sin^2  \frac{ (\lambda_{+} - \lambda_{-} ) L }{4E} 
\nonumber \\
  & &~~ -4 |V_{e+}|^2 |V_{e0}|^2 \sin^2  \frac{ (\lambda_{+} - \lambda_{0} ) L }{4E} 
  \nonumber \\ 
   & &~~ -4 |V_{e0}|^2 |V_{e-}|^2 \sin^2  \frac{ (\lambda_{0} - \lambda_{-} ) L }{4E}
  \nonumber  
\end{eqnarray}
where, L, is the baseline. Now $ |V_{e0}|^2= {\cal O}(\epsilon^2)$, 
so we obtain to order $\epsilon$ eq.~(\ref{eq:P-ee-demo}), or 

\begin{eqnarray}
P(\nu_e \rightarrow \nu_e) &=& 
1 -  \sin^2 2\theta_{13}  ~\left(\frac{\dMsq}{ \lambda_{+} - \lambda_{-} }\right)^2  ~\sin^2 \frac{ (\lambda_{+} - \lambda_{-} ) L }{4E} 
\label{eq:P-ee-sec3}
\end{eqnarray}
where 
$|\lambda_{+} - \lambda_{-}| = \sqrt{  \left(\dMsq - a \right)^2 + 4 s^2_{13} a \dMsq }$ from eq.~(\ref{eq:lambda-pm0}).

Notice that the formula in eq.~(\ref{eq:P-ee-sec3}) takes into account the matter effect as well as the effect of $s_{13}$ to {\it all} orders. Nonetheless, it keeps an exceptional simplicity, 
an effective two-flavor form in matter which consists of single term with the unique eigenvalue difference $\lambda_{+} - \lambda_{-}$, 
the feature we believe to be unique in the market. 
The feature stems from the fact that there is no $\nu_{e}$ component at zeroth order in $\epsilon$ in the ``$0$'' state in matter. It is expressed in the zero in the $V_{e 0}$ element of the zeroth-order $V$ matrix as in (\ref{eq:V0}), see also section~\ref{sec:V-matrix}. 

\subsubsection{$\nu_e \rightarrow \nu_\mu$ and $\nu_e \rightarrow \nu_\tau$ appearance channels}
\label{sec:P-emu}

Now, we discuss the appearance channels $\nu_e \rightarrow \nu_\mu$ and $\nu_e \rightarrow \nu_\tau$. We describe here the simplest way to derive the formulas for the oscillation probabilities starting from the $V$ matrix by using unitarity. The oscillation probability $P(\nu_e \rightarrow \nu_\mu)$ can be computed as 
\begin{eqnarray}
P(\nu_e \rightarrow \nu_\mu) & = & 
\biggl | V_{\mu -} V^*_{e-} e^{ - i \frac{\lambda_{-} L}{2E} } + V_{\mu 0} V^*_{e 0} e^{ - i \frac{\lambda_{0} L }{2E} }  + V_{\mu +} V^*_{e+} e^{ - i \frac{\lambda_{+} L }{2E} } 
\biggr |^2 
\label{P-emu-shortest1}
\end{eqnarray}
We use unitarity relation $V_{\mu -} V^*_{e-} + V_{\mu 0} V^*_{e 0} + V_{\mu +} V^*_{e+}=0$ to eliminate the $V_{\mu -} V^*_{e-}$ term in (\ref{P-emu-shortest1}). Then, we obtain 
\begin{eqnarray}
P(\nu_e \rightarrow \nu_\mu) & = & 4 | V_{\mu +} V^*_{e+} \sin \Delta_{+-} e^{ - i \Delta_{+0}} - V_{\mu 0} V^*_{e0} \sin \Delta_{-0} |^2
\nonumber \\[2mm]
&=& 
4 | V_{\mu+}|^2 |V_{e+}|^2 \sin^2 \Delta_{+-}
\nonumber \\[2mm]
&~&  
- 8 {\cal R}(V_{\mu+} V^*_{e+}V^*_{\mu 0} V_{e0}) \sin \Delta_{+ -} \sin \Delta_{- 0} \cos \Delta_{+ 0} \nonumber \\[2mm]
&~& 
- 8 {\cal I}(V_{\mu+} V^*_{e+}V^*_{\mu 0} V_{e0}) \sin \Delta_{+-} \sin \Delta_{- 0} \sin \Delta_{+ 0}
\nonumber \\[2mm]
&~&
+4|V_{\mu 0}|^2 |V_{e 0}|^2 \sin^2 \Delta_{- 0}
\end{eqnarray}
where the common shorthand notation for the  kinematic phase $\Delta_{i j} = (\lambda_{i}-\lambda_{j})L/4E$ is used. 
Again, since $ |V_{e0}|^2= {\cal O}(\epsilon^2) $, we have to order $\epsilon$
\begin{eqnarray}
&& P(\nu_e \rightarrow \nu_\mu) 
\nonumber \\
&=& 
\left[ 
s^2_{23} \sin^2 2 \theta_{13}
+ 
4 \epsilon   
J_r \cos \delta 
\left\{ \frac{ (\lambda_{+} - \lambda_{-}) - ( \dMsq - \am ) }{  ( \lambda_{+} - \lambda_{0} ) } \right\}
\right]
 \left(\frac{\dMsq}{ \lambda_{+} - \lambda_{-} }\right)^2 \sin^2 \frac{ (\lambda_{+} - \lambda_{-}) L}{ 4E } 
\nonumber \\[2mm]
&+& 8 \epsilon  
J_r 
\frac{ (\dMsq)^3 }{ ( \lambda_{+} - \lambda_{-} ) ( \lambda_{+} - \lambda_{0} ) ( \lambda_{-} - \lambda_{0} ) }
\sin \frac{ (\lambda_{+} - \lambda_{-}) L}{ 4E } 
\sin \frac{ (\lambda_{-} - \lambda_{0}) L}{ 4E} 
\cos \left( \delta - \frac{ (\lambda_{+} - \lambda_{0}) L}{ 4E } \right)
\nonumber \\
\label{eq:P-emu-sec3}
\end{eqnarray}
here $J_r$, the reduced Jarlskog factor, is 
\begin{eqnarray} 
J_r \equiv c_{12} s_{12} c_{23} s_{23} c^2_{13} s_{13}.
\label{eq:Jarlskog-def}
\end{eqnarray}
This expression for the  $\nu_e \rightarrow \nu_\mu$ appearance channel probability is quite compact, despite that it contains all-order contributions of $s_{13}$ and $a$. In particular, it keeps the similar structure as the one derived by the Cervera {\it et al.} 
\cite{Cervera:2000kp}, which retains terms of order $\epsilon^2$ but is expanded by $s_{13}$ only up to second order. 

This method of computing $P(\nu_e \rightarrow \nu_\mu)$ in the above offers the shortest path to the expression of the oscillation probability which is manifestly free from the apparent singularity as $\lambda_{-} \rightarrow  \lambda_{0}$ because $1/(\lambda_{-} - \lambda_{0})$ always appears adjacent to $\sin \left[ (\lambda_{-} - \lambda_{0}) L / 4E  \right]$. We will refer this method as the ``shortcut method'' in the rest of this paper.

The expression $P(\nu_e \rightarrow \nu_\mu)$, eq. (\ref{eq:P-emu-sec3}), is not quite of the form of eq. (\ref{eq:P-ba-general}) because of the $\cos \frac{ (\lambda_{+} - \lambda_{0}) L}{ 4E }$ term in the expansion of  $\cos \left( \delta - \frac{ (\lambda_{+} - \lambda_{0}) L}{ 4E } \right)$. This can be easily remedied, by using the following identity 
\begin{eqnarray}
 2 \sin \Delta_{+-}  \sin \Delta_{-0} \cos \Delta_{+0} & = & \sin^2 \Delta_{+0} - \sin^2 \Delta_{-0} - \sin^2 \Delta_{+-}  
\label{trigonom-identity}
\end{eqnarray} 
which leads to exactly the form of eq. (\ref{eq:P-ba-general}). However, if this is used then some of the terms are singular when $\lambda_{-} = \lambda_{0}$, yet the total  expression is equivalent to eq. (\ref{eq:P-emu-sec3}) and is finite. This is the reason why we prefer the form of eq. (\ref{eq:P-emu-sec3}). 

The oscillation probability for $\nu_e \rightarrow \nu_\tau$ channel can be obtained by all of the following three methods: 
(1) the similar calculation by the shortcut method, 
(2) using the unitarity relation $P( \nu_e \rightarrow \nu_\tau ) = 1- P( \nu_e \rightarrow \nu_e) - P( \nu_e \rightarrow \nu_\mu)$, 
(3) using the relation 
$P(\nu_e \rightarrow \nu_\tau) = P(\nu_e \rightarrow \nu_\mu: c_{23} \rightarrow - s_{23}, s_{23} \rightarrow c_{23})$ \cite{Akhmedov:2004ny}, whose derivation is sketched in section~\ref{sec:basis}. We note that $J_r$ changes sign by the transformation.
In general the oscillation probability in arbitrary channel can be obtained by (1) the shortcut method, or by (2) rewriting the expressions in appendix~\ref{sec:oscillation-P} using the formulas in appendix~\ref{sec:relations}. 

\subsubsection{The general form of the $\nu_{\alpha} \rightarrow \nu_{\beta}$ oscillation probabilities}
\label{sec:P-albe}

The expressions of the oscillation probabilities in the $\nu_{\mu} - \nu_{\tau}$ sector can be derived by one of the methods mentioned in the previous subsection. Then, one observes a remarkable feature that all the oscillation probabilities $P(\nu_\alpha \rightarrow \nu_\beta)$ (including the $\nu_{e}$ sector) can be written in a universal form:
{\small
\begin{eqnarray}
&& P(\nu_\alpha \rightarrow \nu_\beta) =  \delta_{\alpha \beta}
\nonumber \\ 
&& \hspace*{-0.5cm}
+ ~4\left[ 
 \{A^{\alpha \beta}_{+-}\}~s^2_\phi c^2_\phi
+ \epsilon ~\{B^{\alpha \beta}_{+-}\}   \left(J_r \cos \delta \right) 
\frac{ (\dMsq)^2  
\left\{ ( \lambda_{+} - \lambda_{-} ) - ( \dMsq - a ) 
\right\} }{ ( \lambda_{+} - \lambda_{-} )^2 ( \lambda_{+} - \lambda_{0} ) } 
\right]
~\sin^2 \frac{ (\lambda_{+} - \lambda_{-}) L}{ 4E } 
\nonumber \\[2mm] 
&& \hspace*{-0.5cm}
+ 
~4\left[
\{A^{\alpha \beta}_{+0}\}~c_\phi^2 
~+  \epsilon ~\{B^{\alpha \beta}_{+0}\} \left(J_r \cos \delta/c^2_{13} \right) 
\frac{ \dMsq \left\{ ( \lambda_{+} - \lambda_{-} ) - ( \dMsq + a ) \right\} }{ ( \lambda_{+} - \lambda_{-} ) ( \lambda_{+} - \lambda_{0} ) } 
\right]
~\sin^2 \frac{ (\lambda_{+} - \lambda_{0}) L}{ 4E} 
\nonumber \\[2mm] 
&& \hspace*{-0.5cm}
+ 
~4 \left[
\{A^{\alpha \beta}_{-0}\}~ s_\phi^2 
~+\epsilon ~\{B^{\alpha \beta}_{-0}\} \left(J_r \cos \delta/c^2_{13} \right)
\frac{ \dMsq \left\{ ( \lambda_{+} - \lambda_{-} ) + ( \dMsq + a ) \right\} }{ ( \lambda_{+} - \lambda_{-} ) ( \lambda_{-} - \lambda_{0} ) } 
\right]
~\sin^2 \frac{ (\lambda_{-} - \lambda_{0}) L}{ 4E } 
\nonumber\\[2mm]
&& \hspace*{-0.5cm}
 + ~8 \epsilon ~J_r  
~\frac{ (\dMsq)^3 }{ ( \lambda_{+} - \lambda_{-} ) ( \lambda_{+} - \lambda_{0} ) ( \lambda_{-} - \lambda_{0} ) }
~\sin \frac{ (\lambda_{+} - \lambda_{-}) L}{ 4E } 
~\sin \frac{ (\lambda_{-} - \lambda_{0}) L}{ 4E } 
\nonumber \\[2mm]
& & \hspace*{3cm}  \times
\left[
~\{C^{\alpha \beta}\}  ~\cos \delta ~\cos \frac{ (\lambda_{+} - \lambda_{0}) L}{ 4E} 
+ ~\{S^{\alpha \beta}\} ~ \sin \delta ~\sin \frac{ (\lambda_{+} - \lambda_{0}) L}{ 4E } 
~\right], 
\label{eq:P-albe-sec3}
\end{eqnarray}
}
where the eight coefficients $A^{\alpha \beta}_{ij}$,  $B^{\alpha \beta}_{ij}$,  $C^{\alpha \beta}$ and $S^{\alpha \beta}$ are given in Table~\ref{tab:coeff}. Notice that they are $0, ~\pm1$, or the simple functions of $\theta_{23}$. 

\begin{table}[t]
\hspace*{-1.5cm} \begin{tabular}{|c|c|c|c|c|c|c|}
 \hline    & & & & & & \\
 &  $\nu_e \rightarrow  \nu_e$  &     $\begin{array}{c} \nu_e \rightarrow \nu_\mu \\  \nu_\mu \rightarrow \nu_e \end{array}  $  &  
 $\begin{array}{c} \nu_e \rightarrow \nu_\tau \\  \nu_\tau \rightarrow \nu_e \end{array}  $   &  $\begin{array}{c} \nu_\mu \rightarrow \nu_\tau \\  \nu_\tau \rightarrow \nu_\mu \end{array}  $   &  $ \nu_\mu \rightarrow \nu_\mu$  &  $\nu_\tau \rightarrow \nu_\tau $   \\[6mm] \hline
  Order $\epsilon^0$:  & & & & & & \\[1mm]
  $A^{\alpha \beta}_{+-}$ & -1 & $\sin^2 \theta_{23} $ &  $\cos^2 \theta_{23} $ &  $-\sin^2 \theta_{23} \cos^2 \theta_{23} $ &
  $-\sin^4 \theta_{23} $ & $-\cos^4 \theta_{23} $ \\[1mm]
  $A^{\alpha \beta}_{+0} =A^{\alpha \beta}_{-0}$ & 0 & 0 & 0 & $\sin^2 \theta_{23} \cos^2 \theta_{23} $  &  $-\sin^2 \theta_{23} \cos^2 \theta_{23}$ &  $-\sin^2 \theta_{23} \cos^2 \theta_{23}$ 
\\[3mm] \hline  Order $\epsilon \cos \delta$: & & & & & & \\[1mm]
   $B^{\alpha \beta}_{+-} = C^{\alpha \beta}$ &  0 & 1 & -1 & $-\cos 2 \theta_{23}$ &  $\cos 2 \theta_{23}-1$ 
   & $\cos 2 \theta_{23}+1$ \\[1mm] 
   $B^{\alpha \beta}_{+0}=B^{\alpha \beta}_{-0}$  & 0 & 0 & 0 &   $-\cos 2 \theta_{23}$ &  $\cos 2 \theta_{23}$ &  $\cos 2 \theta_{23}$ 
\\[3mm]
\hline    Order $\epsilon \sin \delta$: & & & & & & \\[1mm]
  $  S^{\alpha \beta} $ &  0  & $\pm$1 & $\mp$1 & $\pm$1 & 0 & 0 \\[3mm] \hline
  \end{tabular} 
  \caption{The values for the 5 coefficients for all oscillation channels, $\nu_\alpha \rightarrow \nu_\beta$ and  $\bar{\nu}_\alpha \rightarrow \bar{\nu}_\beta$ to be used in conjunction with eq.~(\ref{eq:P-albe-sec3}). Note that they are $0, ~\pm1$ or simple functions of $\theta_{23}$.}
\label{tab:coeff}
\end{table}

In Table~\ref{tab:coeff}, we observe that three relations hold between the coefficients
\begin{eqnarray}
A^{\alpha \beta}_{-0}= A^{\alpha \beta}_{+0},  \quad B^{\alpha \beta}_{-0}= B^{\alpha \beta}_{+0}  \quad  {\rm and} \quad  B^{\alpha \beta}_{+-}= C^{\alpha \beta}.
\label{eq:ABCids}
\end{eqnarray}
These identities hold due to the invariance of the oscillation probabilities under the following transformation
\begin{equation}
\phi \rightarrow \pi/2 + \phi  \quad  {\rm and} \quad  \lambda_+ \leftrightarrow \lambda_-. 
\label{eq:invar}
\end{equation}
This invariance must be respected by the oscillation probabilities because the two cases in eq. (\ref{eq:invar}) are both equally valid ways of diagonalizing the zeroth-order Hamiltonian. Look at eq.~(\ref{eq:cos-sin-2phi}) to confirm that it is invariant under (\ref{eq:invar}). Then, the former two identities in (\ref{eq:ABCids}) trivially follow, but the last one requires use of the kinematic relationship \footnote{Notice that the relation $B^{\alpha \beta}_{+-}= C^{\alpha \beta}$ needs to be satisfied only to order $\epsilon^0$ because these terms are already suppressed by $\epsilon$. }
\begin{eqnarray}
\sin \Delta_{+-}  \sin \Delta_{+0} \cos \Delta_{-0}  &=& \sin \Delta_{+-}  \sin \Delta_{-0} \cos \Delta_{+0}  +  \sin^2 \Delta_{+-} . 
\nonumber
\end{eqnarray}

We note that $\theta_{23}$ dependence in the oscillation probabilities is confined into the five independent coefficients and all the other parameters, $\Delta m^2$'s, $\theta_{12}$, $\theta_{13}$, $\delta$, $E$ and $a$, are contained in the remaining part of the probabilities, which takes the universal form for all the oscillation channels.  

The antineutrino oscillation probabilities $P(\bar{\nu}_\alpha \rightarrow \bar{\nu}_\beta)$ can be easily obtained from the neutrino oscillation probabilities. One can show that by taking complex conjugate of the evolution equation for anti-neutrinos of energy $E$ that it is equivalent to the evolution equation for neutrino with energy equal to $-E$. 
Therefore, $P(\bar{\nu}_\alpha \rightarrow \bar{\nu}_\beta: E) = P(\nu_\alpha \rightarrow \nu_\beta: - E)$. 

Here, we give some critical observations regarding the formulas (\ref{eq:P-albe-sec3}) and the associated Table~\ref{tab:coeff}: 

\begin{itemize}

\item 

The oscillation probability $P(\nu_e \rightarrow \nu_e)$, to first order in $\epsilon$, is of a simple two-flavor form with the matter-modified mixing angle $\theta_{13}$ and the mass squared difference given by $\dMsq$. As far as we know, having single two-flavor form as the expression of $P(\nu_e \rightarrow \nu_e)$ in matter at this order is the unique case among all the perturbative frameworks so far studied. 

\item 
Formulating the perturbation theory of neutrino oscillation with automatic implementation of the oscillation probability formulas (\ref{eq:P-ba-general}), which is to be carried out in section~\ref{sec:formulation}, is the principal result of this paper. However, the formula (\ref{eq:P-albe-sec3}) goes one step further beyond eq.~(\ref{eq:P-ba-general}) by displaying the channel-independent universal function of $L/E$, up to the coefficients $A_{\pm}$, $B_{\pm}$, etc. which depend only on $\theta_{23}$.

\item 
With the universal form of the probabilities, unitarity can be trivially checked just by adding up the appropriate columns of Table~\ref{tab:coeff} with proper care of using the correct signs in the last row for the intrinsic CP-violating terms.\footnote{ 
Here, we assume that the oscillation probability formulas are derived without using unitarity, excluding one of the options we mentioned earlier to derive them. 
}
One can easily confirm it by adding, for example, the $\nu_e \rightarrow \nu_e$, $\nu_e \rightarrow \nu_\mu$, and $\nu_e \rightarrow \nu_\tau$ columns to give zero in each row. 

\item 
If one uses the identity (\ref{trigonom-identity}) to transform the  
$\sin \frac{ (\lambda_{+} - \lambda_{-}) L}{ 4E } 
\sin \frac{ (\lambda_{-} - \lambda_{0}) L}{ 4E } 
\cos \frac{ (\lambda_{+} - \lambda_{0}) L}{ 4E }$ term to the right-hand side of (\ref{trigonom-identity}), then the transformational invariance given by eq.~(\ref{eq:invar}) will be manifest and eq.~(\ref{eq:P-albe-sec3}) will become the form of eq.~(\ref{eq:P-ba-general}).  However, in the limit $\lambda_{-} \rightarrow \lambda_{0}$ some terms will diverge even thought the total expression is perfectly finite.

\item In the universal form, the amplitude of a given oscillation frequency, to leading (zeroth) order in $\epsilon$, is controlled by $\phi$.
The amplitudes of the frequencies  $(\lambda_{+}-\lambda_{0})$,  $ (\lambda_{-}-\lambda_{0})$ and $ (\lambda_{+}-\lambda_{-})$ are given proportional to $A_{+0} \cos^2 \phi$, $A_{-0}\sin^2\phi$ and $A_{+-}\sin^2 \phi \cos^2 \phi$, respectively.  For the channels involving $\nu_e$, only the $(\lambda_{+}-\lambda_{-})$ frequency appears whereas in the $\nu_\mu - \nu_\tau$ sector all frequencies appear and are of equal magnitude near the $\theta_{13}$ resonance, when $\phi \approx \pi/4$, but every where else either   the frequency $(\lambda_{+}-\lambda_{0}) \approx \dMsq $ or  $(\lambda_{-}-\lambda_{0})\approx \dMsq $ dominates. 

\end{itemize}

Finally, we give more clarifying remarks on generic features of the oscillation probability formulas presented in this section~\ref{sec:results}. 

\begin{itemize} 

\item The term proportional to $\sin \delta$ in $P(\nu_\alpha \rightarrow \nu_\beta)$, in eq.~(\ref{eq:P-albe-sec3}), is equal to 
\begin{eqnarray}
8  \left\{J_r  \sin \delta
\frac{ \epsilon ~(\dMsq)^3 }{ ( \lambda_{+} - \lambda_{-} ) ( \lambda_{+} - \lambda_{0} ) ( \lambda_{-} - \lambda_{0} ) } \right\}
\sin \frac{ (\lambda_{+} - \lambda_{-}) L}{ 4E } 
\sin \frac{ (\lambda_{-} - \lambda_{0}) L}{ 4E } 
 \sin \frac{ (\lambda_{+} - \lambda_{0}) L}{ 4E }   \nonumber
\end{eqnarray}
up to an overall sign. This is because the quantity in $\{ \cdots \}$ is just the Jarlskog factor in matter due to the Naumov-Harrison-Scott identity, \cite{Naumov:1991ju,Harrison:1999df}. 

\item 
Due to the above theorem $\sin \delta$ terms are always accompanied by the Jarlskog factor $J_r = c_{12} s_{12} c_{23} s_{23} c^2_{13} s_{13}$. Though less well known, it can be proven that $\cos \delta$ terms must be proportional to the reduced Jarlskog factor $J_r/c^2_{13}$ \cite{Asano:2011nj}. It can also be shown that both of the $\delta$-dependent terms must come with the suppression factor of $\epsilon$ \cite{Naumov:1991ju,Harrison:1999df,Asano:2011nj}. It is nice to observe that these general properties are realized in our formulas. 

\item 
Association of the factor $\epsilon$ and the all angle factors to the $\delta$-dependent terms indicates that they are genuine three flavor effects. In our general formula (\ref{eq:P-albe-sec3}), all the terms with explicit $\epsilon$ factor are of this $\delta$-dependent terms. Notice that other corrections of order $\epsilon$, though they do not show up explicitly, are implicitly contained in $\dMsq$, $\lambda_{\pm,0}$ and $\phi$ which do not contain any $\delta$ or $\theta_{23}$ dependence.

\item In vacuum, $a=0$, the above oscillation probabilities reproduce the standard results, to first order in $\epsilon$.
The form is somewhat unusual but we have checked that the expressions are identical.  Also, we have checked that in the limit $L/E \rightarrow 0$, where matter effects are negligible, the coefficients of the $(L/E)^n$, for  $n \le 3$, terms in eq. (\ref{eq:P-albe-sec3}), are identical to those in vacuum for all oscillation channels.

\end{itemize}

\begin{figure}
\begin{center}
\vspace{-6mm}
\includegraphics[width=0.48\textwidth]{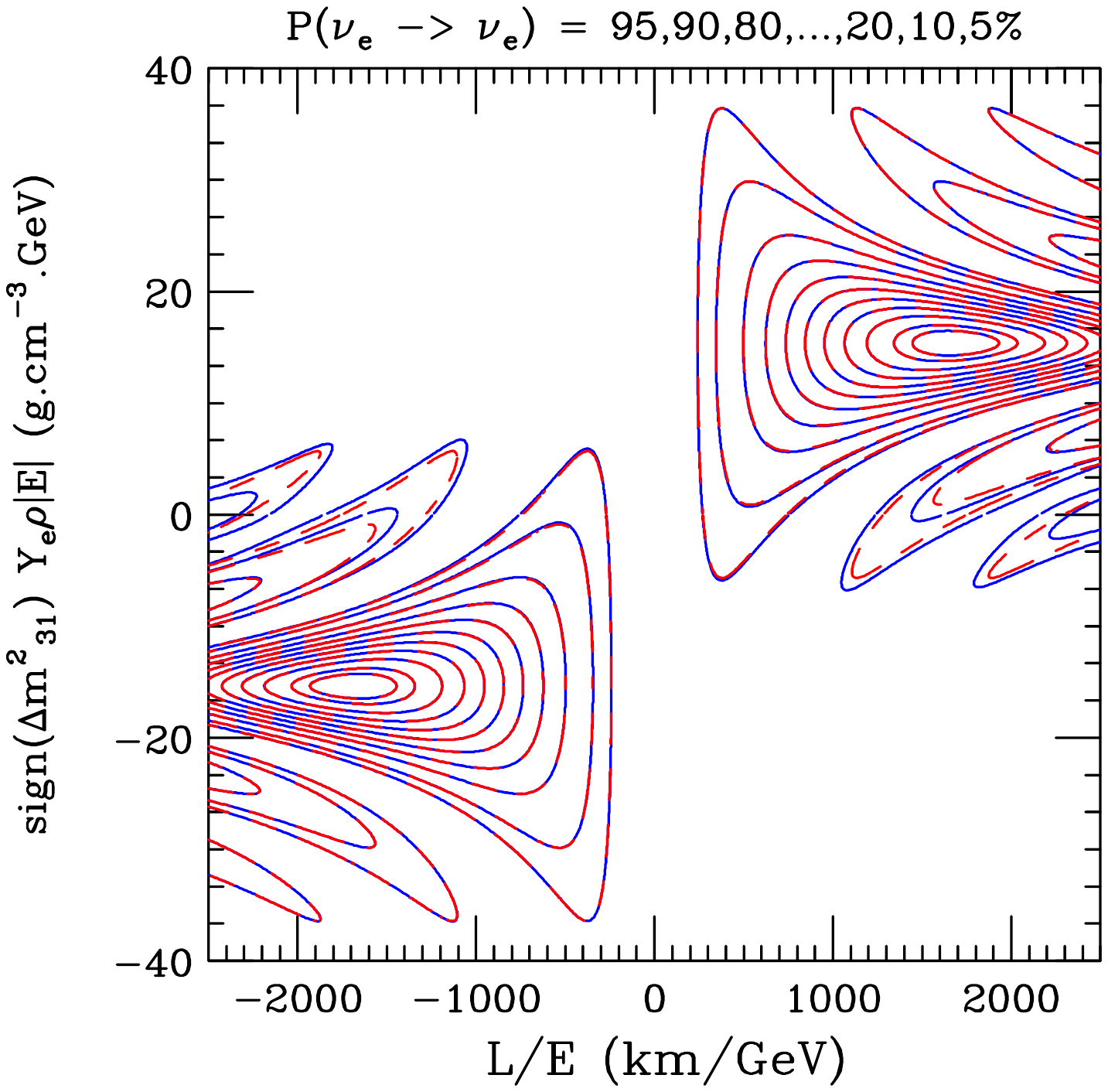}
\includegraphics[width=0.48\textwidth]{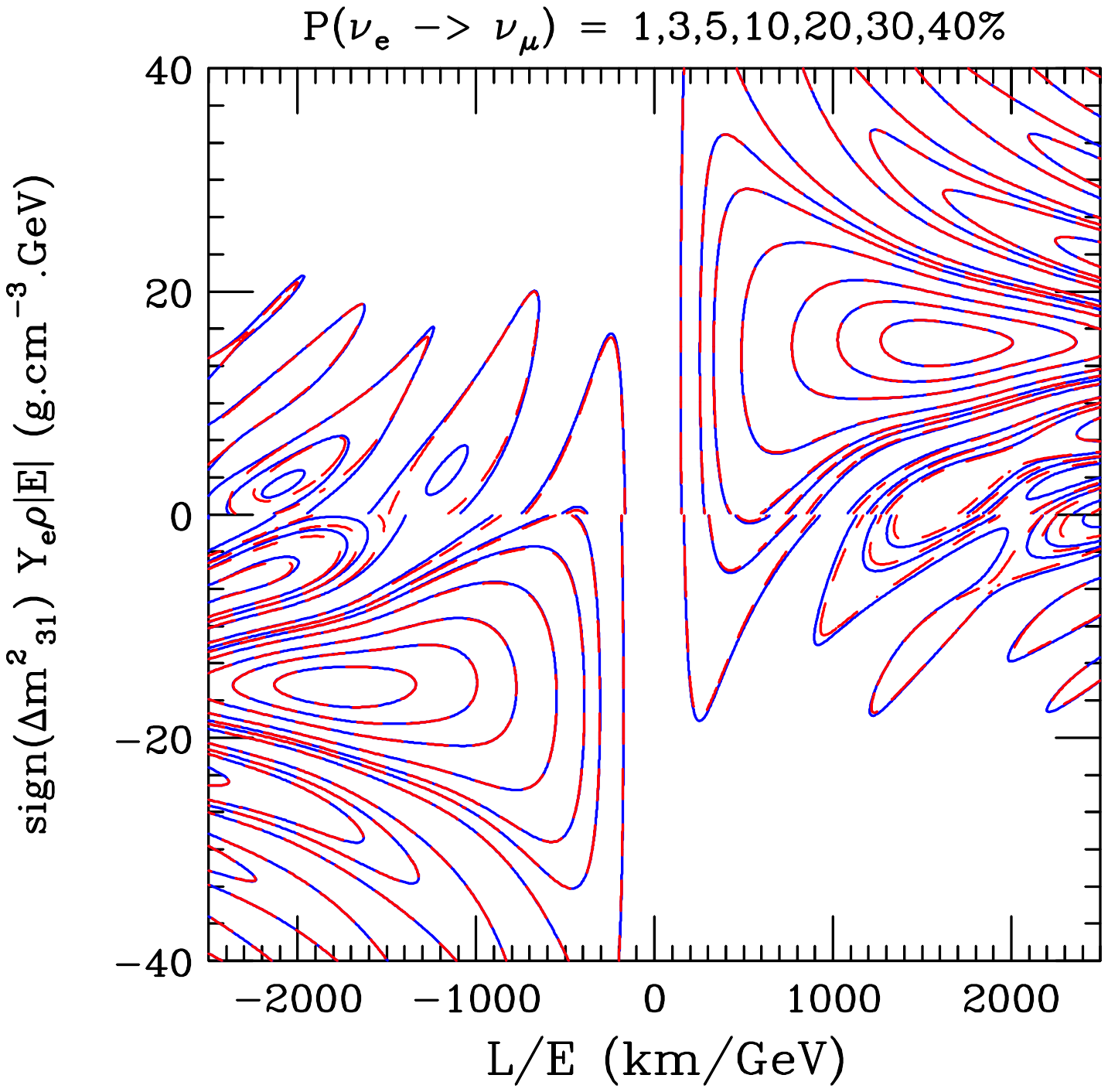}
\includegraphics[width=0.48\textwidth]{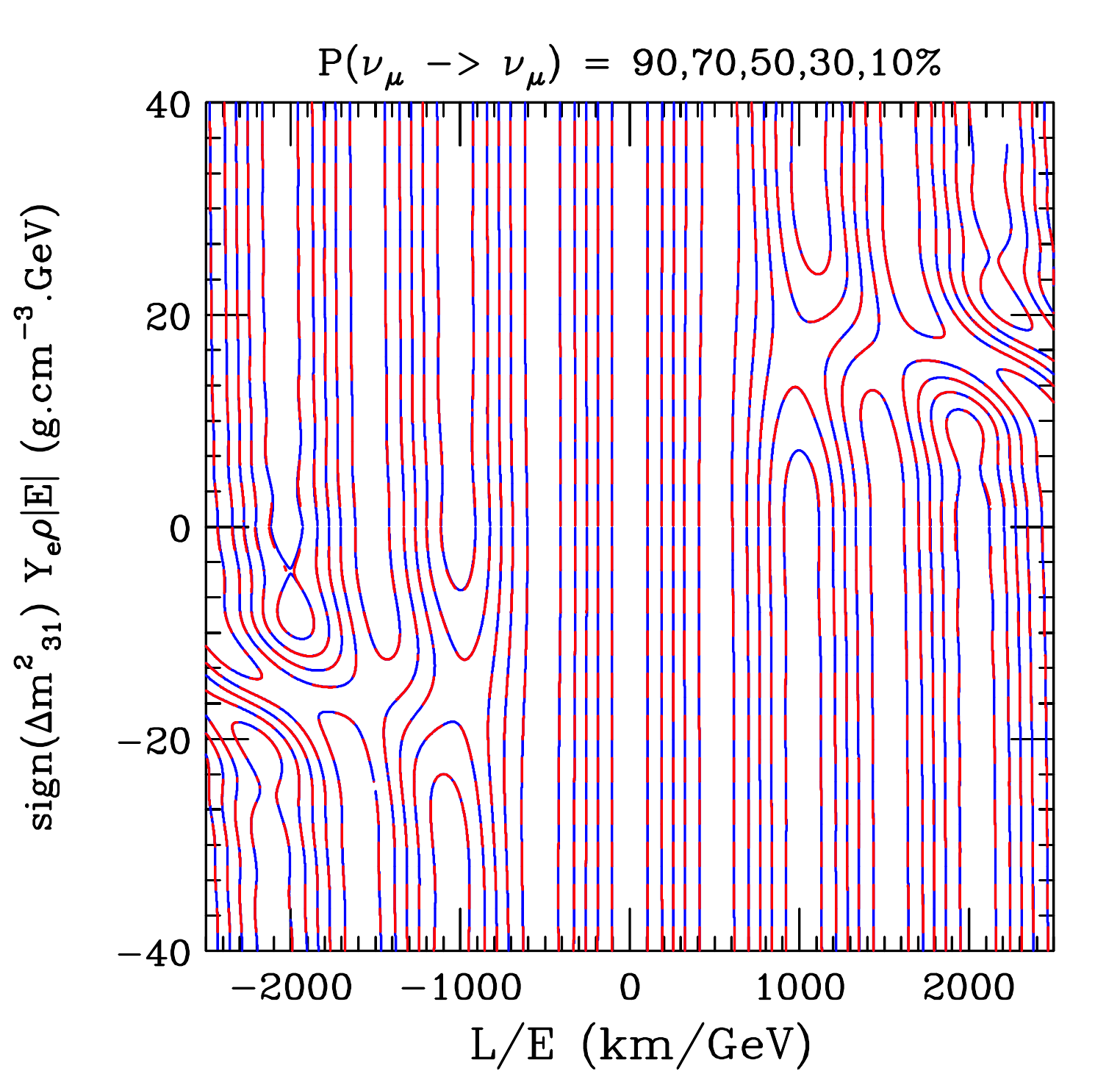}
\end{center}
\vspace{-6mm}
\caption{The iso-probability contours for the exact (solid blue) and approximate (dashed red) oscillation probabilities  for upper left, $\nu_e \rightarrow \nu_e$, upper right, $\nu_e \rightarrow \nu_\mu$ and lower, $\nu_\mu \rightarrow \nu_\mu$.
 The upper (lower) half plane is for normal ordering (inverted ordering), whereas positive (negative) L/E is for neutrinos (antineutrinos). For treatment of antineutrinos, see section~\ref{sec:P-albe}. 
The order of the contours given in the title is determined from the line L/E=0.
The discontinued as one crosses $Y_e \rho |E| =0$ is because we are switching mass orderings at this point.
In most of parameter space the approximate and exact contours sit on top of one another so the lines appear to alternate blue-red dashed. Note that, for L/E  $>$1000 km/GeV and $|Y_e \rho E| < 5$ g.cm$^{-3}$.GeV, the difference between the exact and approximate contours becomes noticeable at least for $\nu_e \rightarrow \nu_e$ and $\nu_e \rightarrow \nu_\mu$.
}
\label{fig:contours}
\end{figure}

\subsubsection{Range of applicability of the formulas}

To discuss the range of applicability of our expressions, it is useful to first consider the vacuum expressions to first order in the expansion parameter $\epsilon$.  For all channels, the expansion of the vacuum oscillation probabilities to first order in  $\epsilon$ does not include terms proportional to $\sin^2 \Delta_{21}$ which starts at second order in $\epsilon$,
\begin{eqnarray}
\sin^2 \Delta_{21} & = & \epsilon^2 \Delta^2_{ren} + {\cal O}( \epsilon^4) \simeq \epsilon^2 \Delta^2_{31} + {\cal O}( \epsilon^4).
\end{eqnarray}
where $\Delta_{ji} \equiv \Delta m^2_{ji} L / 4E $.
When, $\Delta_{31} = \pi/2$, that is at the first atmospheric oscillation maximum, $\epsilon^2 \Delta^2_{31} \approx 0.002$ which is small for the channels $\nu_e \rightarrow \nu_x$ where $x=e, \mu$, or $\tau$ since
$1-P(\nu_e \rightarrow \nu_e)$, $P(\nu_e \rightarrow \nu_\mu)$ and $P(\nu_e \rightarrow \nu_\tau)$ are all of order
$\sin^2 2\theta_{13} \approx 0.1$.  However, at the second atmospheric oscillation maximum, $\Delta^2_{31} =3 \pi/2$ and $\epsilon^2 \Delta^2_{31} \approx 0.02$, which is significant compared to the $\sin^2 2\theta_{13}$ term. So in vacuum our first order expansion is only a good approximation for $\Delta^2_{31} \lsim \pi$ or $L/E \lsim 1000$ km/GeV for these $\nu_e$ channels.
For the other channels, $\nu_\mu \rightarrow \nu_\mu$, $\nu_\tau \rightarrow \nu_\tau$ and $\nu_\mu \rightarrow \nu_\tau$,  our first order expansion is a good approximation to somewhat beyond  L/E $=$ ~1000 km/GeV because the leading terms are not suppressed by the smallness of $\sin^2 2\theta_{13}$.

Then, what about the validity in matter? In section~\ref{sec:more-about} we will argue that our perturbative description is valid outside the solar resonance. Notice that the region without validity (no guarantee for approximation being good) is rather wide and includes the vacuum because the solar resonance width $| \Delta a | = \sqrt{3} (\sin 2 \theta_{12}/ \cos^2 \theta_{13})  \Delta m^2_{21}$  is larger than the solar resonance position $a = (\cos 2 \theta_{12}/ \cos^2 \theta_{13}) \Delta m^2_{21}$. 
We expect then that our helio-perturbation theory works for the matter potential $a$ larger than a few tenth of $| \dMsq| $. 

To give the reader a sense of the precision of our approximation we have plotted in Fig.~\ref{fig:contours}, the contours of equal probability for the exact and the approximate solutions for the channels  $\nu_e \rightarrow \nu_\mu$,  $\nu_e \rightarrow \nu_e$ and  $\nu_\mu \rightarrow \nu_\mu$. The right (left) half plane of each panel of Fig.~\ref{fig:contours} corresponds to the neutrino (anti-neutrino) channel. As expected, for large values of the matter potential,  $|a| > \frac{1}{3}  |\dMsq| $ we find we have no restrictions on  L/E, to have a good approximation to the exact numerical solutions. Whereas for small  values of the matter potential,  $|a| < \frac{1}{3}  |\dMsq| $ we still need the restriction $L/E \lsim 1000$ km/GeV.\footnote{
Of course, the boundary between these two regions should be interpreted as an approximate one.  In fact, an exact boundary would sensitively depend on the definition of the difference allowed between the exact and the approximate probabilities. }

We note that most of the settings for the ongoing and the proposed experiments, except possibly for the one which utilizes the second oscillation maximum, fall into the region $L/E \lsim 1000$ km/GeV. To improve the accuracy to larger values of L/E, especially for values of 
$|a| <  \frac{1}{3}  |\dMsq| $, second order perturbation theory in $\epsilon$ is needed, which will be the subject of a future publication. 

\subsubsection{An example of the power of our oscillation probabilities }
\label{sec:Pee}

\begin{figure}
\begin{center}
\vspace{-10mm}
\includegraphics[width=0.70\textwidth]{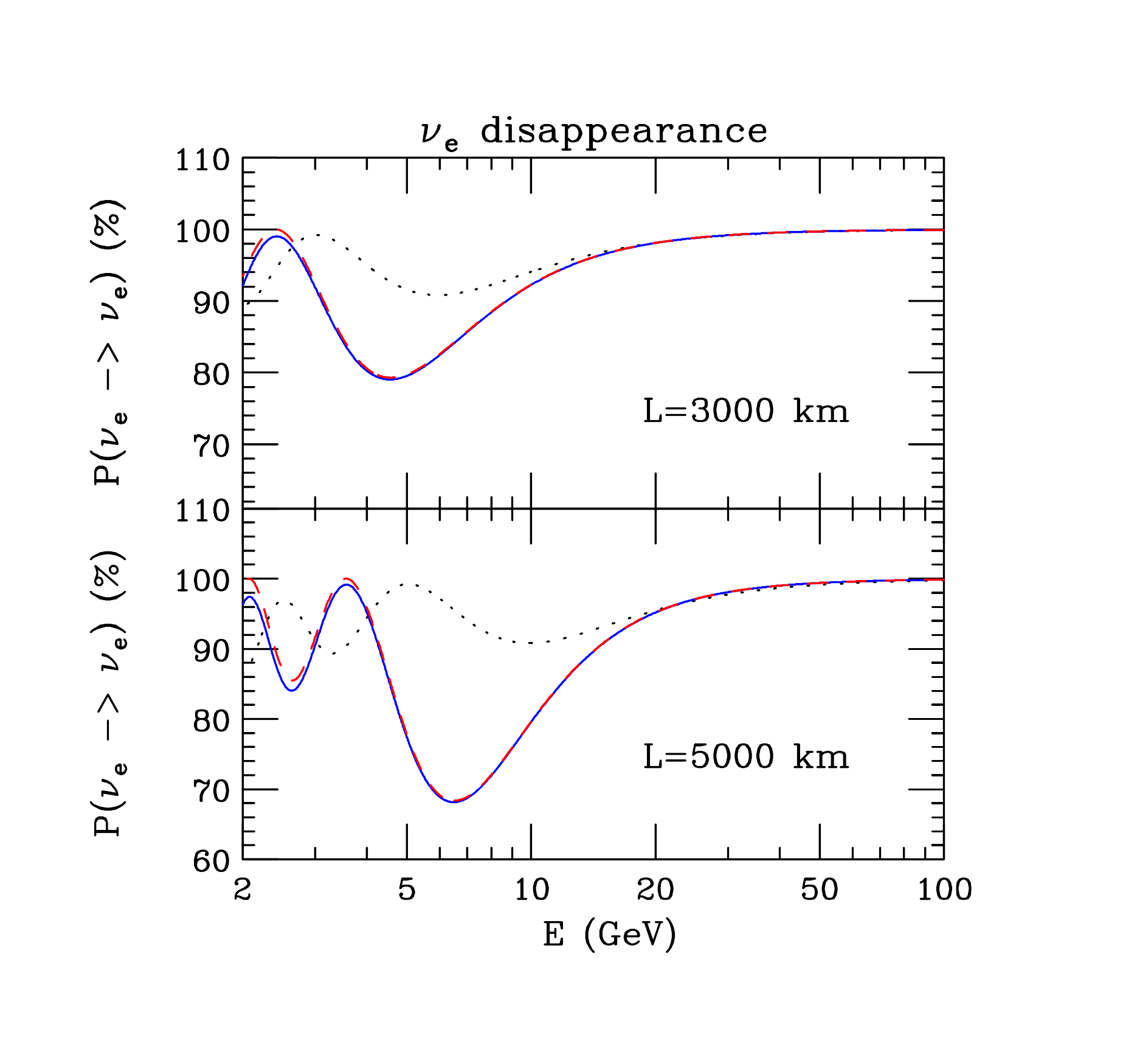}
\end{center}
\vspace{-16mm}
\caption{
The $\nu_{e}$ disappearance probability as a function of energy $E$ for baselines of 3000 km (upper panel) and 5000 km (lower panel). 
We have used the earth matter density 2.8 g/cm$^3$.
}
\label{fig:Pee}
\end{figure}

In this section we present an example of  the power of our compact expression of the oscillation probabilities in understanding the physics of oscillations to first order in our expansion parameter $\epsilon$. For simplicity we consider the $\nu_e$ disappearance channel.   In fig.~\ref{fig:Pee} the $\nu_{e}$ disappearance probability is shown, as a function of neutrino energy $E$, for baselines, $L$, of 3000 km (upper panel) and 5000 km (lower panel). As in figure~\ref{fig:contours}, the solid blue curves are drawn by using the exact expression of $P(\nu_e \rightarrow \nu_e)$, and the dashed red curves by our compact formula in eq.~(\ref{eq:P-ee-sec3}). The black dotted curves are for the vacuum case. 

We first notice that our approximate formula agrees well with the exact expression in particular at higher energy. Secondly, because of the long baselines, the matter effect is sizeable, producing not only a large shift in the position of the first oscillation minimum but also an significant increase in the depth of the minimum.   These two effects are correlated by  the energy dependent quantity, $(\lambda_+ - \lambda_-)$ which can be seen graphically in Fig. \ref{fig:eigenvalue-flow}:  the depth of the oscillation minimum changes from
\begin{eqnarray}
\sin^2 2 \theta_{13} \quad  \rightarrow   \quad \left(  \frac{\dMsq}{\lambda_+ - \lambda_-} \right)^2 \sin^2 2 \theta_{13} \nonumber
\end{eqnarray}
whereas the energy at which the the first oscillation minimum occurs changes from
\begin{eqnarray}
\frac{\dMsq L}{2\pi}  \quad \rightarrow  \quad \frac{(\lambda_+ - \lambda_-)L} {2\pi }.   \nonumber 
 \end{eqnarray}
Because of the simplicity of our expression for  $P(\nu_e \rightarrow \nu_e)$, eq.~(\ref{eq:P-ee-sec3}), these shifts are accurate to first order in the expansion parameter $\epsilon$.  This simple understanding of the features of $P(\nu_e \rightarrow \nu_e)$ is new to this paper.

\subsubsection{Comparison with the existing perturbative frameworks}
\label{sec:comparison}

As we emphasized in section~\ref{sec:structure}, our machinery has an advantage over the existing perturbative frameworks by having the minimum number of terms composed of $\sin \left[ (\lambda_{j} - \lambda_{i}) L / 4E  \right]$ ($i, j = 1, 2, 3$) in the oscillation probabilities. This contrasts with the features of the existing perturbative frameworks in which much larger number of terms than those minimally necessary as in (\ref{eq:P-ba-general}) are produced. They include, typically, the terms with either extra $L/E$ dependences or different frequencies in the sine functions, or often both, which easily obscures the physical interpretation. 

To give a feeling to the readers on how simple and compact our formulas are, we compare our expressions to the ones in the existing literatures to the same order in expansion. For definiteness, we pick the ones in ref.~\cite{Akhmedov:2004ny} to make the comparison, the most recent one among the reference list given in section~\ref{sec:introduction}. 

Our expression of $P(\nu_e \rightarrow \nu_e)$ in (\ref{eq:P-ee-sec3}) which has only a single term (ignoring unity) may be compared with eqs.~(4.6) and (4.7) which consist of total 3 terms. With regard to $P(\nu_e \rightarrow \nu_\mu)$, if we count numbers of terms with different $L/E$ dependence we have one $\delta$-independent and 2 $\delta$-dependent terms in (\ref{eq:P-emu-sec3}), total 3. Whereas eqs.~(4.8) and (4.9) in \cite{Akhmedov:2004ny} have total 7  terms, 3 $\delta$-independent and 4 $\delta$-dependent ones. Our expression of $P(\nu_\mu \rightarrow \nu_\tau)$ in (\ref{eq:P-albe-sec3}) has 3 $\delta$-independent and 2 $\delta$-dependent terms, adding up to total 5. On the other hand, eqs.~(4.10) and (4.11) in \cite{Akhmedov:2004ny} contain total 8 $\delta$-independent and 10 $\delta$-dependent terms (2 $\sin \delta$ and 8 $\cos \delta$ terms), which adds up to 18. So not only do our expressions for the oscillation probabilities have less than half the number of $L/E$ structures, but the form of the $L/E$ dependence is made manifest and identical to the vacuum form. 
%

We note that when our formulas are expanded by $\Delta m^2_{21} / \Delta m^2_{31}$, they agree, of course, with the existing ones calculated previously. Therefore, our formalism may be regarded as a systematic way of organizing the equivalent expressions to order $\Delta m^2_{21} / \Delta m^2_{31}$ into neater, structure-revealing ones. The benefit for having such simple expressions is that the physical interpretation of the formulas is transparent, as emphasized in section~\ref{sec:structure}. 

\section{Formulating the renormalized helio-perturbation theory}
\label{sec:formulation}

In this section, we formulate the helio-to-terrestrial ratio perturbation theory, for short the helio-perturbation theory, which has the unique expansion parameter 
\begin{eqnarray} 
\epsilon \equiv \frac{ \Delta m^2_{21} }{ \dMsq }.
\label{eq:def-epsilon2}
\end{eqnarray}
We will show that use of its renormalized version is the key to the very simple formulas of the oscillation probabilities exhibited in section~\ref{sec:compact-F} and appendix~\ref{sec:oscillation-P}. In fact, there are two ways of deriving the oscillation probabilities, the $S$ matrix method and the wave function method. Here we sketch both of them, leaving technical or computational parts into Appendices~\ref{sec:calculating-S-matrix} and~\ref{sec:oscillation-P}. The meaning of the agreement between results obtained by both the $S$ matrix and the wave function methods will be discussed at the end of this section.

The $S$ matrix describes neutrino flavor changes $\nu_{\beta} \rightarrow \nu_{\alpha}$ after traversing a distance $L$, 
\begin{eqnarray} 
\nu_{\alpha} (L) = S_{\alpha \beta} \nu_{\beta} (0), 
\label{eq:def-S}
\end{eqnarray}
and the oscillation probability is given by 
\begin{eqnarray} 
P(\nu_{\beta} \rightarrow \nu_{\alpha}; L)= 
\vert S_{\alpha \beta} \vert^2.  
\label{eq:def-P}
\end{eqnarray}
When the neutrino evolution is governed by the Schr\"odinger equation, 
$ i \frac{d}{dx} \nu = H \nu  $, 
$S$ matrix is given as 
\begin{eqnarray} 
S = T \text{exp} \left[ -i \int^{L}_{0} dx H(x) \right] 
\label{eq:evolution}
\end{eqnarray}
where $T$ symbol indicates the ``time ordering''  
(in fact ``space ordering'' here). 
In the standard three-flavor neutrinos, Hamiltonian is given by 
\begin{eqnarray}
H= 
\frac{ 1 }{ 2E } \left\{ 
U \left[
\begin{array}{ccc}
0 & 0 & 0 \\
0 & \Delta m^2_{21}& 0 \\
0 & 0 & \Delta m^2_{31} 
\end{array}
\right] U^{\dagger}
+
\left[
\begin{array}{ccc}
a(x) & 0 & 0 \\
0 & 0 & 0 \\
0 & 0 & 0
\end{array}
\right] 
\right\}, 
\label{eq:hamiltonian}
\end{eqnarray}
where the symbols are defined in an earlier footnote. For the case of constant matter density, the right-hand side of (\ref{eq:evolution}) may be written as 
$e^{-i H x}$. 
We recapitulate here the earlier footnote: In (\ref{eq:hamiltonian}) $\Delta m^2_{ji} \equiv m^2_{j} - m^2_{i}$ where $m_{i}$ denotes the mass of $i$-th mass eigenstate neutrinos. Position dependent function $a(x) \equiv 2\sqrt{2} G_F N_e(x) E$ is a coefficient for measuring the matter effect on neutrinos propagating in medium of electron number density 
$N_e(x)$~\cite{Wolfenstein:1977ue} where $G_F$ is the Fermi constant and $E$ is the neutrino energy. 

The neutrino flavor mixing matrix $U$ is usually taken to be the standard form $U_{\text{\tiny PDG}}$ given by Particle Data Group. We, however, prefer to work in a slightly different basis, for this paper, in which the flavor mixing matrix has a form (with the obvious notations $s_{ij} \equiv \sin \theta_{ij}$ etc. and $\delta$ being the CP violating phase)
\begin{eqnarray}
U &=& 
\left[
\begin{array}{ccc}
1 & 0 &  0  \\
0 & 1 & 0 \\
0 & 0 & e^{- i \delta} \\
\end{array}
\right] 
U_{\text{\tiny PDG}} 
\left[
\begin{array}{ccc}
1 & 0 &  0  \\
0 & 1 & 0 \\
0 & 0 & e^{ i \delta} \\
\end{array}
\right] 
\nonumber \\
&=&
\left[
\begin{array}{ccc}
1 & 0 &  0  \\
0 & c_{23} & s_{23} e^{ i \delta} \\
0 & - s_{23} e^{- i \delta} & c_{23} \\
\end{array}
\right] 
\left[
\begin{array}{ccc}
c_{13}  & 0 &  s_{13} \\
0 & 1 & 0 \\
- s_{13} & 0 & c_{13}  \\
\end{array}
\right] 
\left[
\begin{array}{ccc}
c_{12} & s_{12}  &  0  \\
- s_{12} & c_{12} & 0 \\
0 & 0 & 1 \\
\end{array}
\right] 
\equiv 
U_{23} U_{13} U_{12} 
\label{eq:MNS-matrix}
\end{eqnarray}
under the understanding that the left phase matrix in the first line in eq.~(\ref{eq:MNS-matrix}) is to be absorbed into the $\nu_\tau$ neutrino wave functions. Being connected by the phase matrix rotation, it is obvious that our $U$ and $U_{\text{\tiny PDG}}$ give rise to the same neutrino oscillation probabilities. 

\subsection{Choosing the basis for the renormalized helio-perturbation theory }
\label{sec:basis}

It is convenient to work with the tilde basis defined as $\tilde{\nu}_{\alpha} = (U_{23}^{\dagger})_{\alpha \beta} \nu_{\beta}$, in which the Hamiltonian is related to the flavor basis one as 
\begin{eqnarray} 
\tilde{H} = U_{23}^{\dagger} H U_{23}, 
\label{eq:tilde-hamiltonian}
\end{eqnarray}
where $U_{23}$ is defined in eq.~(\ref{eq:MNS-matrix}).
The $S$ matrix in the flavor basis is related to the $S$ matrix in the tilde basis $\tilde{S}$ as
\begin{eqnarray} 
S(L) = U_{23} \tilde{S} (L) U_{23}^{\dagger},
\hspace{10mm}
\tilde{S} (L) = T \text{exp} \left[ -i \int^{L}_{0} dx \tilde{H} (x)  \right].
\label{eq:S-Stilde-matrix}
\end{eqnarray}
Notice that the matter term in the Hamiltonian (\ref{eq:hamiltonian}) is invariant under the $U_{23}$ rotation. Hence, the dynamics of neutrino propagation in matter is governed by only the two mixing angles $\theta_{12}$ and $\theta_{13}$ \cite{Kuo:1989qe}, which is also independent of $\delta$ thanks to our convention of $U$ in (\ref{eq:MNS-matrix}). From the first equation in (\ref{eq:S-Stilde-matrix}), the relationships $P( \nu_e \rightarrow \nu_\tau) = P( \nu_e \rightarrow \nu_\mu: c_{23} \rightarrow - s_{23}, s_{23} \rightarrow c_{23})$ etc. simply follow as used in section~\ref{sec:compact-F}. 

The simplest formulation of helio-perturbative treatment in the tilde-basis includes decomposition of $\tilde{H}$ into the zeroth and the first order terms in the expansion parameter $ \frac{ \Delta m^2_{21} }{ \Delta m^2_{31} }$ as  
\begin{eqnarray} 
\tilde{H} (x) &=& 
\frac{ \Delta m^2_{31} }{2E} 
\left\{
\left[
\begin{array}{ccc}
\frac{ a(x) }{ \Delta m^2_{31} } + s^2_{13} & 0 & c_{13} s_{13} \\
0 & 0 & 0 \\
c_{13} s_{13} & 0 & c^2_{13} 
\end{array}
\right] 
+ \frac{ \Delta m^2_{21} }{ \Delta m^2_{31} }
\left[
\begin{array}{ccc}
s^2_{12} c^2_{13} & c_{12} s_{12} c_{13} & - s^2_{12} c_{13} s_{13} \\
c_{12} s_{12} c_{13} & c^2_{12} & - c_{12} s_{12} s_{13} \\
- s^2_{12} c_{13} s_{13} & - c_{12} s_{12} s_{13} & s^2_{12} s^2_{13} 
\end{array}
\right] 
\right\}
\nonumber \\
\label{eq:H-tilde-usual}
\end{eqnarray} 

To derive the compact formulas of oscillation probabilities we use slightly different basis, a renormalized basis, to formulate the perturbation theory. That is, we absorb a certain order $\epsilon$ terms into the zeroth order Hamiltonian, 
$\tilde{H} (x) = \tilde{H}_{0} (x) + \tilde{H}_{1} (x)$:
\begin{eqnarray} 
\tilde{H}_{0} (x) &=& 
\frac{ \dMsq }{ 2E } \left\{
\left[
\begin{array}{ccc}
\frac{ a(x) }{ \dMsq}  + s^2_{13} & 0 & c_{13} s_{13} \\
0 & 0 & 0 \\
c_{13} s_{13}  & 0 & c^2_{13} 
\end{array}
\right] 
+
\epsilon 
\left[
\begin{array}{ccc}
s^2_{12} & 0 & 0 \\
0 & c^2_{12} & 0 \\
0 & 0 & s^2_{12} 
\end{array}
\right]  \right\}
\label{eq:H-tilde-zeroth}
\\[3mm]
\tilde{H}_{1} (x) &=& 
\epsilon c_{12} s_{12} \frac{ \dMsq }{ 2 E }
\left[
\begin{array}{ccc}
0 & c_{13} & 0 \\
c_{13} & 0 & - s_{13}  \\
0 & - s_{13} & 0 
\end{array}
\right] 
\label{eq:H-tilde-first}
\end{eqnarray} 
where $\dMsq \equiv \Delta m^2_{31} - s^2_{12} \Delta m^2_{21}$ and $\epsilon \equiv \Delta m^2_{21}/\dMsq $, as defined in (\ref{eq:Dm2-ren-def}) and (\ref{eq:epsilon-def}).  $\tilde{H} (x)$ with (\ref{eq:H-tilde-zeroth}) and (\ref{eq:H-tilde-first}) is identical with the tilde-Hamiltonian in (\ref{eq:H-tilde-usual}).  Note the simplicity of the perturbing Hamiltonian, $\tilde{H}_{1}$, especially the zeros on the diagonal elements. The diagonalization of $\tilde{H}_{0}$ leads to the eigenvalues given in section~\ref{sec:lambdas}, eq.~(\ref{eq:lambda-pm0}), and because of the zeros along the diagonal of $\tilde{H}_{1}$, there are no corrections to these eigenvalues at first order in a perturbative expansion. 

The authors of \cite{Blennow:2013rca} treat the order $\epsilon$ effect in the Hamiltonian as a renormalization of the matter potential, whereas we regard it as a renormalization of $\Delta m^2_{31}$.

Our treatment can be easily generalized to cases with matter density variation as far as the adiabatic approximation holds. See e.g., \cite{Akhmedov:2004ny} for such a treatment. However, we prefer to remain, for simplicity and compactness of the expressions, into the formulas with constant matter density approximation in the rest of this paper. 

\subsection{Hat basis}
\label{sec:hat-basis}

We transform the Hamiltonian $\hat{H} = \hat{H}_{0} + \hat{H}_{1}$, from the``tilde'' basis to the ``hat'' basis, using 
\begin{eqnarray} 
\hat{H}_{0} &=& U^{\dagger}_{\phi} \tilde{H}_{0} U_{\phi}, 
\hspace{10mm}
\hat{H}_{1} = U^{\dagger}_{\phi} \tilde{H}_{1} U_{\phi} 
\label{eq:hat-hamiltonian1}
\end{eqnarray}
where the unperturbed Hamiltonian $\hat{H}_{0}$ is diagonal, 
\begin{eqnarray} 
\hat{H}_{0} &=& 
\frac{1}{ 2E }
\left[
\begin{array}{ccc}
\lambda_{-} & 0 & 0 \\
0 & \lambda_{0} & 0 \\
0 & 0 & \lambda_{+}
\end{array}
\right]. 
\label{eq:H0-hat}
\end{eqnarray}
We take the following form of unitary matrix $U_{\phi}$ to diagonalize $\tilde{H}_{0}$:
\begin{eqnarray} 
U_{\phi} =
\left[
\begin{array}{ccc}
\cos \phi & 0 & \sin \phi \\
0 & 1 & 0 \\
- \sin \phi & 0 & \cos \phi 
\end{array}
\right].  
\label{eq:U-phi}
\end{eqnarray}
The expressions of the zeroth order eigenvalues $\lambda_{-}$, $\lambda_{0}$, and $\lambda_{+}$, are given in 
eqn (\ref{eq:lambda-pm0}). Similarly, cosine and sine $\phi$ are given in eqn. (\ref{eq:cos-sin-2phi}). 

Also the perturbing Hamiltonian, $\hat{H}_{1}$, retains it's simple form thanks to that the $U_{\phi}$ rotation keeps ``zeros'' in $\tilde{H}_{1}$ unchanged both on the diagonal and in the top-right and bottom-left corners, 
\begin{eqnarray} 
&& \hat{H}_{1} = U^{\dagger}_{\phi} \tilde{H}_{1} U_{\phi} 
\nonumber\\
&=&
\epsilon c_{12} s_{12} \frac{ \Delta m^2_{\rm ren} }{ 2 E }
\left[
\begin{array}{ccc}
0 & \cos \left( \phi - \theta_{13} \right) & 0   \\
\cos \left( \phi - \theta_{13} \right)  & 0 & \sin \left( \phi - \theta_{13} \right) \\
0  & \sin \left( \phi - \theta_{13} \right) & 0 
\end{array}
\right]. 
\label{eq:H1-hat}
\end{eqnarray}
In fact, $ \hat{H}_{1}$ is identical to $\tilde{H}_1$ with $\theta_{13}$ replaced by $(\theta_{13}-\phi)$.

\subsection{S Matrix and the oscillation probability }
\label{sec:S-matrix}

The $S$ matrix in the flavor basis is related to the $S$ matrix in the tilde and the hat bases as 
\begin{eqnarray} 
S(L) = U_{23} \tilde{S} (L) U_{23}^{\dagger} = U_{23} U_{\phi} \hat{S} (L) U^{\dagger}_{\phi} U_{23}^{\dagger} 
\label{eq:S-Stilde-Shat-matrix}
\end{eqnarray}
where we have used explicitly the fact that the matter density is constant: 
\begin{eqnarray} 
\tilde{S} (L) = T \text{exp} \left[ -i \int^{L}_{0} dx \tilde{H} (x)  \right] = U_{\phi} T \exp \left[ -i \int^{L}_{0} dx \hat{H} (x)  \right] U^{\dagger}_{\phi} \equiv U_{\phi} \hat{S} (L) U^{\dagger}_{\phi}.
\label{eq:tilde-Smatrix}
\end{eqnarray}
To calculate $\hat {S} (L)$ we define $\Omega(L)$ as 
\begin{eqnarray} 
\Omega(L) = e^{i \hat{H}_{0} L} \hat{S} (L).
\label{eq:def-omega}
\end{eqnarray}
Then, $\Omega(L)$ obeys the evolution equation 
\begin{eqnarray} 
i \frac{d}{dx} \Omega(x) = \check{H}_{1} \Omega(x) 
\label{eq:omega-evolution}
\end{eqnarray}
where 
\begin{eqnarray} 
\check{H}_{1} \equiv e^{i \hat{H}_{0} x} \hat{H}_{1} e^{-i \hat{H}_{0} x} .
\label{eq:def-H1}
\end{eqnarray}
Then, $\Omega(x)$ can be computed perturbatively as 
\begin{eqnarray} 
\Omega(L) = 1 + 
(-i) \int^{L}_{0} dx \check{H}_{1} (x) + 
\mathcal{O} ( \epsilon^2 ). 
\label{eq:Omega-exp}
\end{eqnarray}
Collecting the formulas the $S$ matrix can be written as 
\begin{eqnarray} 
S(L) =  
U_{23} U_{\phi} e^{-i \hat{H}_{0} L} \Omega(L) U^{\dagger}_{\phi} U_{23}^{\dagger} 
\label{eq:hat-Smatrix}
\end{eqnarray}
Thus, we are left with perturbative computation of $\Omega(L)$ with use of (\ref{eq:def-H1}) to calculate the $S$ matrix. With the $S$ matrix in hand it is straightforward to compute the oscillation probabilities by using (\ref{eq:def-P}). We leave these tasks to appendices~\ref{sec:calculating-S-matrix} and \ref{sec:oscillation-P}. 

\subsection{Mass eigenstate in matter: $V$ matrix method}
\label{sec:V-matrix}

In this section we calculate the $V$-matrix directly using our perturbation theory.
If we switch off the perturbation $\hat{H}_{1}$, the mass eigenstates in matter, to lowest order, are given by the hat-basis wave function $\hat{\nu}_{i}^{(0)}$, which are the eigenstates of $\hat{H}_{0}$ in (\ref{eq:hat-hamiltonian1}), and since $\hat{H}_{0}$
is diagonal, we have
\begin{eqnarray}
\hat{\nu}_{i}^{(0)} = (U_{23} U_{\phi})^\dagger_{i \alpha}  \nu_{\alpha}.
\label{eq:flavor-matter-0th}
\end{eqnarray}
Thus, the $V$ matrix is given to zeroth order by $V^{(0)} = U_{23} U_\phi$ whose explicit form is given in section~\ref{sec:V-matter}, and also in eq.~(\ref{eq:V0}).

In order to obtain the mass eigenstates in matter to first order in $\epsilon$, $\nu_{i} = \hat{\nu}_{i}^{(0)} + \hat{\nu}_{i}^{(1)}$, let us compute the first order correction to the hat basis wave functions. Using the familiar perturbative formula for the perturbed wave functions 
\begin{eqnarray}
\hat{\nu}_{i}^{(1)} = 
2E~ \sum_{j\neq i} \frac{ (\hat{H}_{1})_{ji} }{ \lambda_i - \lambda_j }
~\hat{\nu}_{j}^{(0)}
\label{eq:nu-hat-first-order}
\end{eqnarray}
with $\hat{H}_{1}$ in (\ref{eq:hat-hamiltonian1}), and the $\lambda_i$'s are given by the eigenvalues of $\hat{H}_{0}$,
see (\ref{eq:lambda-pm0}).
Then the mass eigenstate in matter $\nu_{i}$ can be written to first order in $\epsilon$ as:

\begin{eqnarray}
\left( \begin{array}{c} \nu_- \\ \nu_0 \\ \nu_+ \end{array} \right) & = & 
\left(
\begin{array}{ccc}
1 &  \epsilon \dMsq \frac{ c_{12} s_{12} c_{ \left( \phi - \theta_{13} \right) } }{\lambda_{-} - \lambda_{0} }  & 0   \\
- \epsilon \dMsq \frac{ c_{12} s_{12} c_{ \left( \phi - \theta_{13} \right) } }{\lambda_{-} - \lambda_{0} } & 1 
& -\epsilon \dMsq 
\frac{ c_{12} s_{12} s_{ \left( \phi - \theta_{13} \right) } }{\lambda_{+} - \lambda_{0} } \\
0  & \epsilon \dMsq 
\frac{ c_{12} s_{12} s_{ \left( \phi - \theta_{13} \right) } }{\lambda_{+} - \lambda_{0} } & 1 
\end{array}
\right)
\left( \begin{array}{c} \hat{\nu}_{-}^{(0)}  \\ \hat{\nu}_{0}^{(0)}  \\ \hat{\nu}_{+}^{(0)}  \end{array} \right).
\nonumber 
\end{eqnarray}
Using (\ref{eq:flavor-matter-0th}), this equation is of the form $\nu_i = V^\dagger \nu_\alpha$ which can be inverted to easily obtained the $V$-matrix given in eq.~(\ref{eq:V}), 
\begin{eqnarray}
V & = & U_{23} U_\phi  
\left[ 
\mathds{1}+\epsilon c_{12} s_{12} \dMsq  \left\{ \frac{  c_{ \left( \phi - \theta_{13} \right) } }{\lambda_{-} - \lambda_{0} }  \left( \begin{array}{ccc}
0 & -1 & 0 \\
1 & 0 & 0 \\
0 & 0 & 0
\end{array}
\right)
+  \frac{  s_{ \left( \phi - \theta_{13} \right) } }{\lambda_{+} - \lambda_{0} }   \left( \begin{array}{ccc}
0 & 0 & 0 \\
0 & 0 & 1 \\
0 & -1 & 0
\end{array}
\right) \right\}
\right].
\nonumber \\  & & 
\label{eq:Vfactorized}
\end{eqnarray}
This can be used to directly compute the oscillation probabilities as was performed in section (\ref{sec:compact-F}), or by inserting the $V$ matrix elements into the expressions of the oscillation probabilities in (\ref{eq:MNS-matrix}). 
This $V$ matrix method was used to calculate the matter effect correction in the oscillation probabilities \cite{Minakata:1998bf}. For a different approach toward simpler expressions of the $V$ matrix elements see \cite{Agarwalla:2013tza}.

In sections~\ref{sec:S-matrix} and \ref{sec:V-matrix}, we have sketched the two different methods for calculating the oscillation probabilities, the $S$-matrix method (section~\ref{sec:S-matrix}) and the $V$-matrix method (section~\ref{sec:V-matrix}). The results obtained by the both methods agree with each other, and the expressions of the oscillation probabilities are of the form given in eq.~(\ref{eq:P-ba-general}). 

Let us make a comment on what happens if we use a different decomposition of the Hamiltonian into unperturbed and perturbed parts as in (\ref{eq:H-tilde-usual}). Then, the perturbed Hamiltonian has nonzero diagonal terms, and this produces first order corrections to the eigenvalues. If we use these expressions for the eigenvalues and expand by the small expansion parameter we obtain terms that do not exist in (\ref{eq:P-ba-general}), such as the ones proportional to $\sin \left[ (\lambda_{j} - \lambda_{i}) L / 2E  \right]$. Of course, the $S$-matrix method will give the same expressions. Therefore, absence of the explicit first-order correction to the eigenvalues is essential to keep perturbative expressions of the oscillation probabilities of the form (\ref{eq:P-ba-general}) to order $\epsilon$. In our case it is guaranteed by vanishing diagonal elements of the perturbed Hamiltonian (\ref{eq:H-tilde-first}). 

\section{More about the renormalized helio-perturbation theory}
\label{sec:more-about}

In this section, we critically examine the framework of the renormalized helio-perturbation theory. Despite a drawback of the current framework (which is to be described below) we argue that our perturbation theory works apart from the vicinity of the solar resonance crossing.

\begin{figure}
\begin{center}
\vspace{-6mm}
\includegraphics[width=0.48\textwidth]{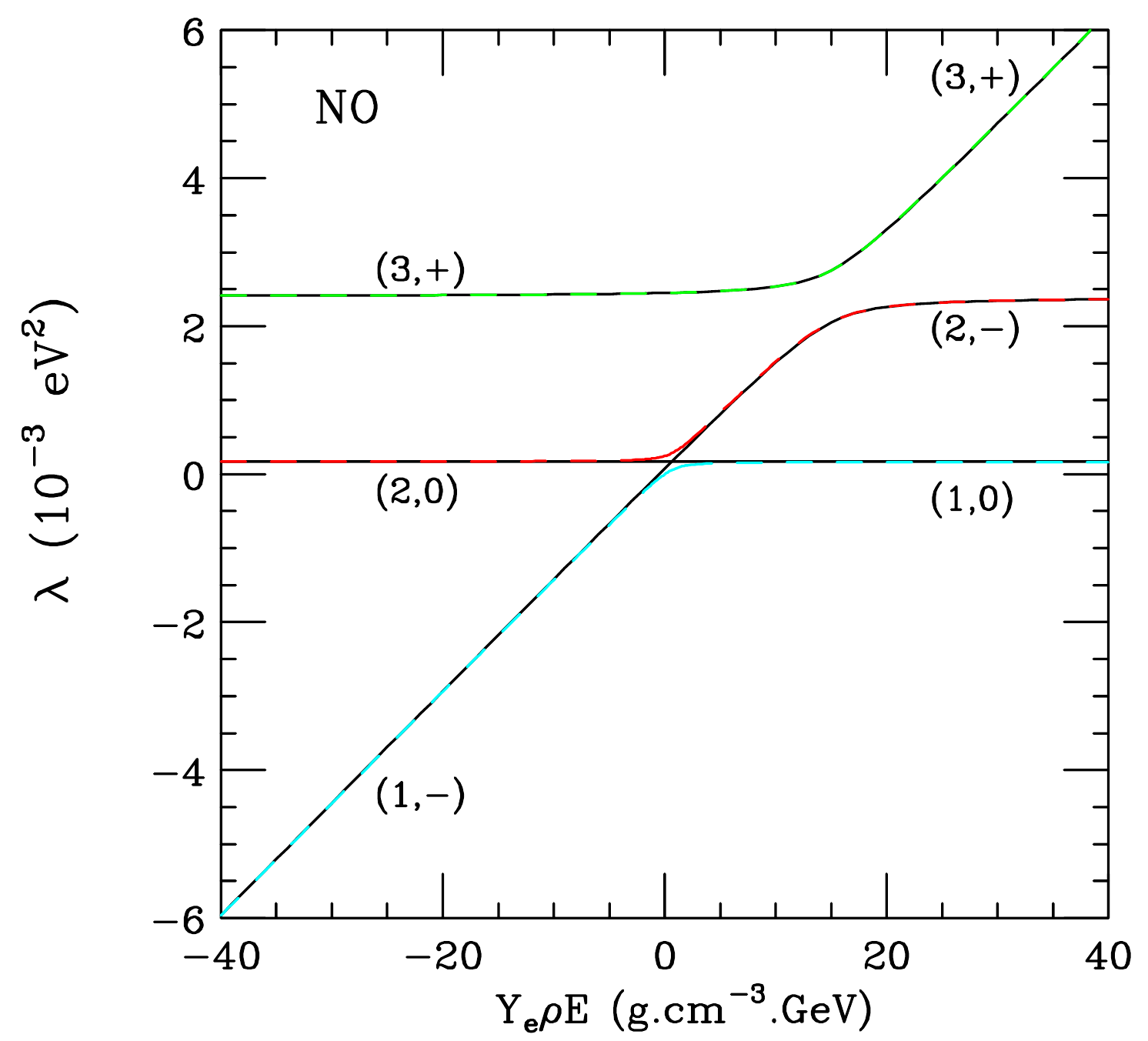}
\includegraphics[width=0.48\textwidth]{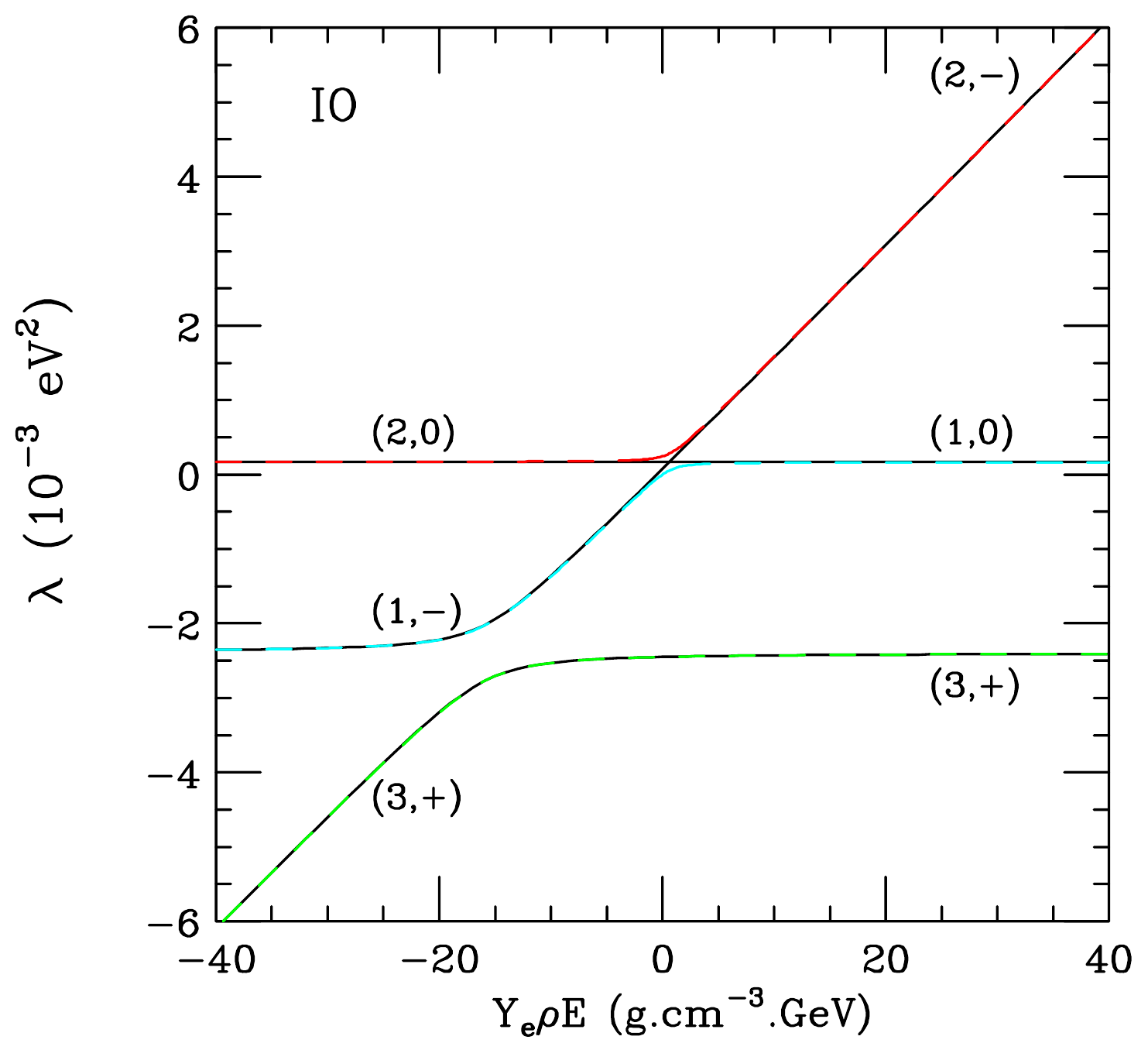}
\end{center}
\vspace{-6mm}
\caption{
The flow of the three eigenvalues in the normal mass ordering, abbreviated as NO, (left panel) and in the inverted mass ordering, abbreviated as IO, (right panel). The exact eigenvalues are depicted with colored lines, green for $\nu_{1}$, red for $\nu_{2}$, and flight blue for $\nu_{3}$. The eigenvalues calculated by using our renormalized helio-perturbation theory are drawn by the black dashed lines whose state labels are marked on the figures.  The approximate eigenvalues for $\nu_{0}$ and  $\nu_{-}$ cross at the solar resonance, whereas the exact eigenvalues for $\nu_{1}$ and $\nu_{2}$ repelled at solar resonance. To make these  features visible we used a solar $\Delta m^{2}_{21}$ three times as large as the measured value.
}
\label{fig:eigenvalue-flow}
\end{figure}

\subsection{Exact versus zeroth-order eigenvalues in matter}
\label{sec:exact-perturbative}

The three eigenvalues of the Hamiltonian are written as $\frac{\lambda_{i}}{2E}$, where $i$ runs over $1,2,3$ for the exact eigenvalues, and $i=-, 0, +$ for the zeroth-order eigenvalues in our perturbative framework. In Figure~\ref{fig:eigenvalue-flow} the $\lambda_{i}$ are plotted as a function of $a$, the Wolfenstein matter potential, for both the exact and the zeroth-order ones given by eq.~(\ref{eq:lambda-pm0}). It is clear in Fig.~\ref{fig:eigenvalue-flow} that our zeroth-order eigenvalues fail to treat the solar-$\Delta m^2$ scale level crossing correctly. As one goes through the solar resonance in the exact solution the two eigenvalues involved, the red and green, repel one another, whereas in our perturbative solution the two corresponding eigenvalues cross with each other.

We point out here that the drawback is the common feature among the similar perturbative treatments of neutrino oscillation available in the market. In this section we argue, for the first time, that despite the drawback, our perturbative framework successfully treat the flavor content of these two states in region reasonably far from the solar resonance. 

We first note that, despite the feature, the atmospheric and solar resonances occur at the correct values of the matter potential with our zeroth-order eigenvalues. The atmospheric resonance occurs when $\lambda_{+}- \lambda_{-}$ is a minimum, which is at 
\begin{equation}
a = \dMsq \cos 2 \theta_{13}  \nonumber
\end{equation}
and the minimum difference between $\lambda_{+}$ and $\lambda_{-}$ is $\dMsq \sin 2 \theta_{13}$ as expected. The solar resonance occurs when $\lambda_{-}- \lambda_{0} $ is a minimum; this occurs when 
\begin{equation}
a =\epsilon \dMsq \cos 2 \theta_{12}/ \cos^2 \theta_{13} 
\nonumber
\end{equation}
and the minimum difference is zero!  This is not the value determined by the full Hamiltonian which is $ \approx \epsilon \dMsq \sin 2 \theta_{12}$.    Therefore, while our perturbative scheme treats the atmospheric resonance correctly to order $\epsilon$, it misses the effects of the solar resonance. 

\subsection{Eigenvalues in vacuum and in the asymptotic regions $a \rightarrow \pm \infty$}
\label{sec:asymptotics}

In this and the next subsections, we will give the arguments to indicate that our renormalized helio-perturbation theory works apart from the vicinity of the solar resonance despite the issue mentioned above. 

We first show that the zeroth-order eigenvalues given in (\ref{eq:lambda-pm0}) agrees with the exact ones to order $\epsilon$ in the asymptotic regions $a \rightarrow \pm \infty$. We use the characteristic equation of the full Hamiltonian 
(\ref{eq:hamiltonian}) to derive ($i=1,2,3$)
\begin{eqnarray}
\sum_{i} \lambda_{i} &=& \left( a + \Delta m^2_{31} + \Delta m^2_{21} \right), 
\nonumber \\
\sum_{i, j} \lambda_{i} \lambda_{j} &=& \Delta m^2_{21} \Delta m^2_{31} + a \left\{ (c^2_{12} + s^2_{12} s^2_{13} ) \Delta m^2_{21} + c^2_{13} \Delta m^2_{31}  \right\}, 
\nonumber \\
\lambda_{1} \lambda_{2} \lambda_{3} &=& c^2_{12} c^2_{13} a \Delta m^2_{21} \Delta m^2_{31}. 
\label{eq:lambda-eq}
\end{eqnarray}
Then, by using the asymptotic expansion of $\lambda$'s one can obtain to leading order in the $1/a$ expansion (at $a \rightarrow + \infty$ in the NO case) 
\begin{eqnarray}
\lambda_{2} = & c^2_{13} \Delta m^2_{31} + s^2_{12} s^2_{13} \Delta m^2_{21} &=(c^2_{13} + s^2_{12} \epsilon ) \Delta m^2_{\rm ren}, 
\nonumber \\
\lambda_{1} = & c^2_{12} \Delta m^2_{21} &=c^2_{12} \epsilon \Delta m^2_{\rm ren}, 
\nonumber \\
\lambda_{3} = & a + s^2_{13} \Delta m^2_{31} + s^2_{12} c^2_{13} \Delta m^2_{21}  &=a + (s^2_{13} +  s^2_{12} \epsilon) \Delta m^2_{\rm ren} .
\label{eqn:L-asymptotic}
\end{eqnarray}
At $a \rightarrow - \infty$, $\lambda_{+}$ and $\lambda_{-}$ must be interchanged. They are identical to the ones computed with our zeroth-order eigenvalues given in (\ref{eq:lambda-pm0}):
\begin{eqnarray}
\left( \begin{array}{c}
\lambda_{-} \\ \lambda_{0} \\ \lambda_{+}
\end{array} \right)_{a=+\infty}^{NO}
= 
\left( \begin{array}{c}
(c^2_{13} + s^2_{12} \epsilon ) \Delta m^2_{\rm ren}   \\ 
c^2_{12} \epsilon \Delta m^2_{\rm ren} \\ 
a + (s^2_{13} +  s^2_{12} \epsilon) \Delta m^2_{\rm ren} 
\end{array} \right)
\label{eq:lambda-pm-asymptotic}
\end{eqnarray}
\begin{eqnarray}
\left( \begin{array}{c}
\lambda_{-} \\ \lambda_{0} \\ \lambda_{+}
\end{array} \right)_{a=-\infty}^{NO}
=
\left( \begin{array}{c}
a + (s^2_{13} +  s^2_{12} \epsilon) \Delta m^2_{\rm ren} \\ 
c^2_{12} \epsilon ~\Delta m^2_{\rm ren} \\ 
(c^2_{13} + s^2_{12} \epsilon ) \Delta m^2_{\rm ren}  
\end{array} \right)
\label{eq:lambda-mp-asymptotic}
\end{eqnarray}
For the IO, the analytic expressions of $\lambda_{i}$ ($i = -, 0, +$) at $a \rightarrow \pm \infty$ is the same as those of $\lambda_{i}$ at $a \rightarrow \mp \infty$ for the case of NO.\footnote{
In fact, the first line in (\ref{eq:lambda-eq}) holds exactly.
}

When the matter potential vanishes, $a =0$, the mass squared eigenvalues are given for the both mass orderings by
\begin{eqnarray}
\left( \begin{array}{c}
\lambda_{-} \\ \lambda_{0} \\ \lambda_{+}
\end{array} \right)_{a=0}^{NO, IO}
=
\left( \begin{array}{c}
s^2_{12} \epsilon \Delta m^2_{\rm ren}   \\ 
c^2_{12} \epsilon \Delta m^2_{\rm ren} \\ 
\Delta m^2_{31}
\end{array} \right)
\label{eq:lambda-vac}
\end{eqnarray}
whose order $\epsilon$ terms are different from the exact values, $\lambda_{1} = 0$, $\lambda_{2} = \Delta m^2_{21}$, and $\lambda_{3} = \Delta m^2_{31}$. To understand the meaning of the failure at order $\epsilon$ we need to discuss what happens to the flavor content of the matter mass eigenstates as the matter potential changes from below to above the solar resonance. This will be done in the next subsection.

Similarly, the asymptotic behaviour of the angle $\phi$, i.e., $\theta_{13}$ in matter can be easily worked out. With the definition of $\phi$ in (\ref{eq:cos-sin-2phi}), it is easy to show that $\phi$ takes on the following values as $a$ is varied from $-~\infty$ to $+ ~ \infty$. In the case of normal mass ordering, 
\begin{eqnarray}
\phi_{NO} & = & 
\left\{ \begin{array}{cl}
0, & a= -~\infty \\
\theta_{13}, &  a= 0 \\
\frac{\pi}{4}, & a =  \dMsq \cos 2 \theta_{13} \\
~~\frac{\pi}{2} - \theta_{13}, ~~ & a=  2 \dMsq \cos 2 \theta_{13} \\
\frac{\pi}{2}, &  a= +~\infty.
\end{array} \right.
\end{eqnarray}
while in the case of inverted mass ordering, 
\begin{eqnarray}
\phi_{IO} & = & 
\left\{ \begin{array}{cl}
\frac{\pi}{2}, & a= -~\infty \\
~\frac{\pi}{2} - \theta_{13}, ~~ & a=  2 \dMsq \cos 2 \theta_{13} \\
\frac{\pi}{4}, & a = \dMsq \cos 2 \theta_{13} \\
\theta_{13} &  a= 0 \\
0, &  a= +~\infty.
\end{array} \right.
\label{phi-flow}
\end{eqnarray}
It reflects a natural view that physics at $a \rightarrow + \infty$ for the normal mass ordering corresponds to the one at $a \rightarrow - \infty$ for the inverted mass ordering at least to leading order in $\epsilon$. 

\subsection{Neutrino mixing matrix in matter and the $\nu$ flavor content at $a \rightarrow \pm \infty$}
\label{sec:V-matter}

Since the two levels cross at the solar resonance in our perturbative treatment, one may expect that our treatment fails completely beyond the solar resonance, i.e., in the region with matter density higher than the resonance. However, we will show in this subsection that the flavor contents of the three eigenstates are correctly reproduced at least in the asymptotic region. That is, the zeroth-order $V$ matrix describes correctly the asymptotic behaviour of the exact eigenstates in matter. 

Suppose we denote the exact flavor mixing matrix in matter as the (almost standard) $U$ matrix defined in (\ref{eq:MNS-matrix}). Let us discuss the case of NO first. 
At $a \rightarrow - \infty$ $U$ is nothing but $V^{(0)}$ in (\ref{eq:V0}). Notice that $s_{23}$ in matter is frozen to its vacuum value for $\epsilon < 0.1$, and $s_{12} \simeq 0$ at $a \rightarrow - \infty$ \cite{Zaglauer:1988gz}. 
At $a \rightarrow + \infty$, $s_{12} \simeq 1$ and $c_{12} \simeq 0$ the matter $U$ matrix is identical to (\ref{eq:V0}) if we interchange $\nu_{1}$ and $\nu_{2}$ apart from re-phasing factor $-1$ for the new second mass eigenstate.

To make the meaning of this feature clearer we write down here the flavor content of the states $\nu_{i}$ ($i=1,2,3$) at $a \rightarrow \pm \infty$. Noticing that $\nu_{i} = (V^{\dagger})_{i \alpha} \nu_{\alpha}$, the flavor composition at $a \rightarrow - \infty$ is given at zeroth order by 
\begin{eqnarray}
\nu_{1} &=& c_{\phi} \nu_{e} - s_{\phi}(s_{23} e^{ - i \delta} \nu_{\mu} + c_{23}  \nu_{\tau}) = \nu_{-}
\nonumber \\
\nu_{2} &=& \{ c_{23} \nu_{\mu} - s_{23} e^{ i \delta} \nu_{\tau} \} =  \nu_{0}
\nonumber \\
\nu_{3} &=& s_{\phi} \nu_{e} + c_{\phi} (s_{23}  e^{ - i \delta} \nu_{\mu} + c_{23}  \nu_{\tau} ) = \nu_{+}
\label{eq:F-content-infty}
\end{eqnarray}

Whereas the composition at $a \rightarrow + \infty$ is given by 
\begin{eqnarray}
\nu_{1} &=& - \{ c_{23} \nu_{\mu} - s_{23} e^{ i \delta} \nu_{\tau} \} = - \nu_{0}
\nonumber \\
\nu_{2} &=& c_{\phi} \nu_{e} - s_{\phi} (s_{23}  e^{ - i \delta} \nu_{\mu} + c_{23}  \nu_{\tau} ) = \nu_{-}
\nonumber \\
\nu_{3} &=& s_{\phi} \nu_{e} + c_{\phi} ( s_{23}  e^{ - i \delta} \nu_{\mu} + c_{23}  \nu_{\tau} ) = \nu_{+}
\label{eq:F-content+infty}
\end{eqnarray}%
The flavor compositions given in (\ref{eq:F-content-infty}) and (\ref{eq:F-content+infty}) imply that the flavor content of the lower two mass eigenstate in matter is correctly described in our perturbative framework despite the failure of describing the solar level crossing.   Note, the combinations of $\nu_\mu$ and $\nu_\tau$ in the $(\cdots)$ and $\{\cdots\}$ are orthogonal in the above expressions.

In the case of IO, essentially the same discussion goes through. The asymptotic behavior of $\theta_{12}$ at $a \rightarrow \pm \infty$ is the same as that in NO. The asymptotic behavior of $\theta_{13}$ at $a \rightarrow \pm \infty$ for NO is mapped into the one at $a \rightarrow \mp \infty$ for IO. But, this is already taken care of by the definition of $\phi$ given in (\ref{eq:cos-sin-2phi}). Therefore, the same $U$ matrix in matter as the one in NO are obtained at $a \rightarrow \pm \infty$. Then, the same flavor compositions as in (\ref{eq:F-content-infty}) and (\ref{eq:F-content+infty}) follow  for IO. Again it is consistent with those we expect from the level crossing diagram. 

We add that our $\nu_{+}$ state always corresponds to $\nu_{3}$ state. At $a \rightarrow + \infty$ in NO and at $a \rightarrow - \infty$ in IO the electron neutrino component is all in this $\nu_{+} = \nu_{3}$ state. At $a \rightarrow - \infty$ in NO and at $a \rightarrow + \infty$ in IO the electron neutrino component is all in $\nu_{-}$ ($\nu_{1}$ for NO, $\nu_{2}$ for IO) state. In vacuum for both NO and IO, $V_{e+}^{(0)} = s_{\phi} = s_{13}$, which is the correct value for the $\nu_{3} = \nu_{+}$ state.

To summarize: 
Despite that the eigenvalues calculated by our helio-perturbation theory do not show the correct behavior at around the solar level crossing, the flavor composition of the states are correctly represented by our zeroth-order states. It is the very reason why our perturbative formulas work at much higher and lower values of $Y_e \rho E$ compared to the one at the solar resonance. This  point escaped detection in the previous literature to our knowledge. We have seen that they work practically whole regions except for the vicinity of the solar resonance. 

\section{Concluding remarks}
\label{sec:conclusion}

We have developed a new perturbative framework of neutrino oscillations, which allows us to derive compact expressions of the oscillation probabilities in matter, which we examined in this paper to order $\epsilon \equiv \Delta m^2_{21} / \dMsq  \simeq \Delta m^2_{21} / \Delta m^2_{31} $. 
Although vacuum like in their forms, our compact results keep all-order effects of $s_{13}$ and the matter potential $a$, while using $\epsilon$ as the unique expansion parameter. 

The characteristic features of our perturbative framework (dubbed as the renormalized helio-perturbation theory), in contrast to the ones previously studied by various authors, are as follows:

\begin{itemize}

\item 

We use the renormalized basis (\ref{eq:H-tilde-zeroth}) defined with the atmospheric mass-squared difference corrected by an order $\epsilon$ quantity, $\dMsq \equiv \Delta m^2_{31} - s^2_{12} \Delta m^2_{21}$. The decomposition of the Hamiltonian into the unperturbed (\ref{eq:H-tilde-zeroth}) and perturbed terms (\ref{eq:H-tilde-first}), is done in such a way that there is no diagonal entries in the perturbed one. It makes the eigenvalues of the zeroth order Hamiltonian the correct ones to order  $\epsilon$. Usage of this decomposition is crucial to obtain the form of the oscillation probabilities (\ref{eq:P-ba-general}) akin to the form in vacuum. This facilitates an immediate physical interpretation that the frequencies of the oscillations are determined by $(\lambda_+-\lambda_-)$, $(\lambda_+-\lambda_0)$, and $(\lambda_--\lambda_0)$, where the $\lambda$'s are the eigenvalues of the zeroth order Hamiltonian, see eq.~(\ref{eq:lambda-pm0}). They determine the oscillation pattern, the functional form of $L/E$ dependence, for all the oscillation probabilities as in eq.~(\ref{eq:P-ba-general}) to this order. 

\item 
The amplitudes of the $\sin \left[ (\lambda_{j} - \lambda_{i}) L / 4E  \right]$ terms  in the oscillation probabilities, (\ref{eq:P-ba-general}), can be determined either by the $S$-matrix method, or the $V$-matrix method. The latter may be simpler because of no first order correction to the eigenvalues. In contrast to the complexity of the exact solution, this perturbative approach ensures calculability of the $V$ matrix elements in the form of power series, see eq.~(\ref{eq:Vfactorized}). It leads to the simple analytic expressions of the oscillation probabilities which have a universal form to first order in $\epsilon$, as shown in eq.~(\ref{eq:P-albe-sec3}). The coefficients in this universal form are $0, \pm1$ or simple functions of $\theta_{23}$, see table \ref{tab:coeff}.

\item
From the universal form of the oscillation probabilities, eq.~(\ref{eq:P-albe-sec3}), we readily observe the followings: \\
\hspace*{0.2cm} 
(1) Each one of the zeroth-order terms in the oscillation probabilities, in all channels, takes the same form as the corresponding two-flavor oscillation probability in vacuum but with use of the eigenvalues and $\theta_{13}$ in matter. In the special case of $\nu_e \rightarrow \nu_e$, only the term with the particular frequency $\propto \lambda_{+} - \lambda_{-}$ is non-zero, so the $L/E$ dependence is of the two flavor form. The dominance of this frequency also occurs in  the oscillation probability for $\nu_e \rightarrow \nu_\mu$ and $\nu_e \rightarrow \nu_\tau$.
\\
\hspace*{0.2cm} 
(2) All the first-order correction terms with the explicit factor $\epsilon$ ($\approx \Delta m^2_{\odot} / \Delta m^2_{\oplus}$) in the oscillation probabilities are proportional to either $\cos \delta$ or $\sin \delta$. They are also suppressed by the angle factor $\propto c_{12} s_{12} c_{23} s_{23} s_{13}$. Hence, they are the genuine three-flavor effects. 
The explicit expressions of the $\delta$-dependent terms in our formulas are in agreement with the general theorems. 
\\
\hspace*{0.2cm} 
(3) Unitarity can be trivially checked. \\
Further comments are made in section~\ref{sec:P-albe}.

\end{itemize}

Despite the success of our current framework in almost all regions including the one around the first oscillation maximum, it has a clear drawback. It fails to accommodate the physics at around the solar-scale resonance. This is related to the unphysical feature that the two eigenvalues ($\lambda_{-}$ and $\lambda_{0}$) cross with each other at the solar resonance, the universal fault in all the perturbative treatment which include the expansion parameter $\epsilon$ in the market. However, we have offered, for the first time to our knowledge, an explanation why our framework works beyond the resonance despite the failure in treating the solar level crossing. We hope that we can return to the problem as a whole in the future. 

As we noticed at the end of section~\ref{sec:lambdas} our renormalized $\dMsq$ is identical, to order $\epsilon$, to the effective $\Delta m^2$ measurable in an (anti-) $\nu_{e}$ disappearance experiment in vacuum. It is a tantalizing question whether it is just a coincidence, or is an indication of something deep. 

It is quite possible that the features of the current framework mentioned above naturally generalizes to higher orders. That is, one can demand that the oscillation probabilities of the general form in (\ref{eq:P-ba-general}) calculated by the $V$-matrix method, with the carefully chosen zeroth-order eigenvalues, agree with the ones from the $S$-matrix method and are correct to certain order in $\epsilon$. Our result in this paper may be regarded as an existence proof of this concept to order $\epsilon$. Since we know that this is true in the exact form of the oscillation probabilities (assuming adiabaticity) \cite{Zaglauer:1988gz}, it is likely to be correct in each order in perturbation theory. It may or may not require higher order renormalization in $\dMsq$.

\appendix

\section{Calculation of $S$ matrix elements}
\label{sec:calculating-S-matrix}

\subsection{Computation of $\hat{S}$ matrix elements}

By using $\hat{H}_{1}$ in (\ref{eq:H1-hat}) and $e^{i \hat{H}_{0} x} = \text{diag} \left[ e^{i \lambda_{-} x/2E}, e^{i \lambda_{0} x/2E}, e^{i \lambda_{+} x/2E} \right]$, one can easily compute $\check{H}_{1}(x) \equiv e^{i \hat{H}_{0} x} \hat{H}_{1} e^{-i \hat{H}_{0} x}$. Then, using eq.~(\ref{eq:Omega-exp}), the first order term of $\Omega(x)$ can be calculated as 
\begin{eqnarray} 
&& \Omega_{1}(x) = 
(-i) \int^{x}_{0} dx' \check{H}_{1} (x') 
\nonumber \\
&=& -i \epsilon \dMsq c_{12} s_{12} 
\left[
\begin{array}{ccc}
0 & c_{\left( \phi - \theta_{13} \right)} \frac{ e^{ i ( \lambda_{-} - \lambda_{0} ) x/2E} - 1 }{ i ( \lambda_{-} - \lambda_{0} )  }, & 0 \\ 
c_{\left( \phi - \theta_{13} \right)} \frac{ e^{ - i ( \lambda_{-} - \lambda_{0} ) x/2E} - 1 }{ - i ( \lambda_{-} - \lambda_{0} )  },
& 0 & s_{\left( \phi - \theta_{13} \right)} \frac{ e^{ - i ( \lambda_{+} - \lambda_{0} ) x/2E} - 1 }{ - i ( \lambda_{+} - \lambda_{0} ) } \\
0 & s_{\left( \phi - \theta_{13} \right)} \frac{ e^{ i ( \lambda_{+} - \lambda_{0} ) x/2E} - 1 }{ i ( \lambda_{+} - \lambda_{0} ) }, 
& 0 
\end{array}
\right] 
\nonumber \\
\label{eq:Omega-1st}
\end{eqnarray}
where we have introduced the simplified notations, $c_{\left( \phi - \theta_{13} \right)} \equiv \cos \left( \phi - \theta_{13} \right)$, $s_{\left( \phi - \theta_{13} \right)} \equiv \sin \left( \phi - \theta_{13} \right)$, etc.
The simplicity in the structure of (\ref{eq:Omega-1st}) with many zeros is the mathematical reason why the expressions of neutrino oscillation probabilities are so simple in our renormalized helio-perturbation theory.

The $\hat{S}$ matrix is given by $\hat{S} = e^{-i \hat{H}_{0} x} \Omega = e^{-i \hat{H}_{0} x} \left[ {\bf 1} + \Omega_{1} (x) \right]$ where {\bf 1} denotes the unit matrix. Then, the elements of $\hat{S}$ matrix are given by: 
\begin{eqnarray} 
\hat{S}_{ee} &=& e^{- i \lambda_{-} x/2E}, 
\hspace{10mm}
\hat{S}_{\mu \mu} = e^{- i \lambda_{0} x/2E},
\hspace{10mm}
\hat{S}_{\tau \tau} = e^{- i \lambda_{+} x/2E}, 
\nonumber \\
\hat{S}_{e \tau} &=& \hat{S}_{\tau e} = 0, 
\nonumber \\ 
\hat{S}_{e \mu} &=& \hat{S}_{\mu e} = 
-i \epsilon \dMsq c_{12} s_{12} c_{\left( \phi - \theta_{13} \right)} \frac{ \left( e^{- i \lambda_{0} x/2E} - e^{- i \lambda_{-} x/2E} \right) }{ i ( \lambda_{-} - \lambda_{0} ) }, 
\nonumber \\
\hat{S}_{\mu \tau} &=& \hat{S}_{\tau \mu} = 
-i \epsilon \dMsq c_{12} s_{12} s_{\left( \phi - \theta_{13} \right)} \frac{ \left( e^{- i \lambda_{0} x/2E} - e^{- i \lambda_{+} x/2E} \right) }{ i ( \lambda_{+} - \lambda_{0} ) }.
\label{eq:hatS-elements}
\end{eqnarray}

\subsection{The relationships between $\hat{S}$, $\tilde{S}$, and $S$ matrices}

The relationships between $\hat{S}$, $\tilde{S}$, and $S$ matrices are summarized as 
\begin{eqnarray} 
\tilde{S} (x) = U_{\phi} \hat{S} (x) U^{\dagger}_{\phi} 
\hspace{10mm}
S(x) = U_{23} \tilde{S} (x) U_{23}^{\dagger} .
\label{eq:hatS-tildeS-S}
\end{eqnarray}
The left equation can be written explicitly as 
 \begin{eqnarray} 
\tilde{S} \equiv 
\left[
\begin{array}{ccc}
\tilde{S}_{e e} & \tilde{S}_{e \mu} & \tilde{S}_{e \tau} \\
\tilde{S}_{\mu e} & \tilde{S}_{\mu \mu} & \tilde{S}_{\mu \tau} \\
\tilde{S}_{\tau e} & \tilde{S}_{\tau \mu} & \tilde{S}_{\tau \tau}
\end{array}
\right] 
&=& 
\left[
\begin{array}{ccc}
c_{\phi} & 0 & s_{\phi} \\
0 & 1 & 0 \\
- s_{\phi} & 0 & c_{\phi} 
\end{array}
\right] 
\left[
\begin{array}{ccc}
\hat{S}_{ee} & \hat{S}_{e \mu} & \hat{S}_{e \tau} \\
\hat{S}_{\mu e} & \hat{S}_{\mu \mu} & \hat{S}_{\mu \tau} \\
\hat{S}_{\tau e} & \hat{S}_{\tau \mu} & \hat{S}_{\tau \tau}
\end{array}
\right] 
\left[
\begin{array}{ccc}
c_{\phi} & 0 & - s_{\phi} \\
0 & 1 & 0 \\
s_{\phi} & 0 & c_{\phi} 
\end{array}
\right] .
\label{eq:tildeS-hatS}
\end{eqnarray}
Hence, the elements of $\tilde{S}$ can by written in terms of the elements of $ \hat{S}$ as 
\begin{eqnarray} 
\tilde{S}_{ee} &=& c^2_{\phi} \hat{S}_{ee} + s^2_{\phi} \hat{S}_{\tau \tau} + c_{\phi} s_{\phi} \left( \hat{S}_{e \tau} + \hat{S}_{\tau e} \right), 
\nonumber \\
\tilde{S}_{e \mu} &=& c_{\phi} \hat{S}_{e \mu} + s_{\phi} \hat{S}_{\tau \mu},
\nonumber \\ 
\tilde{S}_{e \tau} &=& c^2_{\phi} \hat{S}_{e \tau} - s^2_{\phi} \hat{S}_{\tau e} - c_{\phi} s_{\phi} \left( \hat{S}_{e e} - \hat{S}_{\tau \tau} \right), 
\nonumber \\
\tilde{S}_{\mu e} &=& c_{\phi} \hat{S}_{\mu e} + s_{\phi} \hat{S}_{\mu \tau} = \tilde{S}_{e \mu} 
\nonumber \\
\tilde{S}_{\mu \mu} &=& \hat{S}_{\mu \mu} 
\nonumber \\
\tilde{S}_{\mu \tau} &=& - s_{\phi} \hat{S}_{\mu e} + c_{\phi} \hat{S}_{\mu \tau},
\nonumber \\ 
\tilde{S}_{\tau e} &=& - s^2_{\phi} \hat{S}_{e \tau} + c^2_{\phi} \hat{S}_{\tau e} - c_{\phi} s_{\phi} \left( \hat{S}_{e e} - \hat{S}_{\tau \tau} \right) = \tilde{S}_{e \tau},
\nonumber \\ 
\tilde{S}_{\tau \mu} &=& - s_{\phi} \hat{S}_{e \mu} + c_{\phi} \hat{S}_{\tau \mu} = \tilde{S}_{\mu \tau}  
\nonumber \\
\tilde{S}_{\tau \tau} &=& s^2_{\phi} \hat{S}_{ee} + c^2_{\phi} \hat{S}_{\tau \tau} - c_{\phi} s_{\phi} \left( \hat{S}_{e \tau} + \hat{S}_{\tau e} \right). 
\label{eq:tildeS-elements}
\end{eqnarray}
Similarly, the elements of $S$ can be written in terms of  the elements of $\tilde{S}$ matrix as
\begin{eqnarray} 
S_{ee} &=& \tilde{ S }_{ee}, 
\nonumber \\
S_{e \mu} &=& c_{23} \tilde{ S }_{e \mu} + s_{23} e^{ - i \delta} \tilde{ S }_{e \tau}, 
\nonumber \\ 
S_{e \tau} &=& c_{23} \tilde{ S }_{e \tau} - s_{23} e^{ i \delta} \tilde{ S }_{e \mu},
\nonumber \\
S_{\mu e} &=& c_{23} \tilde{ S }_{\mu e} + s_{23} e^{ i \delta} \tilde{ S }_{\tau e} 
= S_{e \mu} (- \delta),  
\nonumber \\
S_{\mu \mu} &=& c^2_{23} \tilde{ S }_{\mu \mu} + s^2_{23} \tilde{ S }_{\tau \tau} + c_{23} s_{23} ( e^{ - i \delta} \tilde{ S }_{\mu \tau} + e^{ i \delta}  \tilde{ S }_{\tau \mu} ), 
\nonumber \\
S_{\mu \tau} &=& c^2_{23} \tilde{ S }_{\mu \tau} - s^2_{23} e^{ 2 i \delta}  \tilde{ S }_{\tau \mu} + c_{23} s_{23} e^{ i \delta}  ( \tilde{ S }_{\tau \tau} - \tilde{ S }_{\mu \mu} ), 
\nonumber \\ 
S_{\tau e} &=& c_{23} \tilde{ S }_{\tau e} - s_{23} e^{ - i \delta} \tilde{ S }_{\mu e} 
= S_{e \tau} (- \delta), 
\nonumber \\ 
S_{\tau \mu} &=& c^2_{23} \tilde{ S }_{\tau \mu } - s^2_{23} e^{ - 2 i \delta} \tilde{ S }_{ \mu \tau} + c_{23} s_{23} e^{ - i \delta} ( \tilde{ S }_{\tau \tau} - \tilde{ S }_{\mu \mu} ) = S_{\mu \tau} (- \delta),
\nonumber \\
S_{\tau \tau} &=& s^2_{23} \tilde{ S }_{\mu \mu} + c^2_{23} \tilde{ S }_{\tau \tau} - c_{23} s_{23} ( e^{ - i \delta} \tilde{ S }_{\mu \tau} + e^{ i \delta} \tilde{ S }_{\tau \mu} ).
\label{eq:S-elements}
\end{eqnarray}
Then, the oscillation probabilities are simply given by $P(\nu_{\beta} \rightarrow \nu_{\alpha}; L) = \vert S_{\alpha \beta} \vert^2 $, eq. (\ref{eq:def-P}).

\section{Expressions of neutrino oscillation probabilities}
\label{sec:oscillation-P}

With the expressions of $S$ matrix elements obtained in appendix~\ref{sec:calculating-S-matrix}, and using eq. (\ref{eq:def-P}) it is straightforward to calculate the neutrino oscillation probabilities. Similarly, one can insert the $V$ matrix elements given in (\ref{eq:V}) into (\ref{eq:P-ba-general}) to obtain the equivalent results. In this appendix we only give the resulting expressions, but of all the oscillation channels under the viewpoint that unitarity is not to be imposed but must be proven to show the consistency of the calculation. To compare these with the results of section (\ref{sec:compact-F}), use of the identities given in Appendix (\ref{sec:relations}) may be of some help. 

The only comment worth to give here is about the simple method for transformation $c_{23} \rightarrow - s_{23}$ and $s_{23} \rightarrow c_{23}$ to obtain $P( \nu_e \rightarrow \nu_\tau)$ from $P( \nu_e \rightarrow \nu_\mu)$, or $P( \nu_\tau \rightarrow \nu_\tau)$ from $P( \nu_\mu \rightarrow \nu_\mu)$, 
which is utilized in section~\ref{sec:results}.  
Though we work with the rephased flavor mixing matrix defined in (\ref{eq:MNS-matrix}), the transformations produces $S_{e \tau}$ from $S_{e \mu}$, and $S_{\tau \tau}$ from $S_{\mu \mu}$, up to an overall phase, see eq. (\ref{eq:S-elements}).

\subsection{Oscillation probabilities in $\nu_e  -$row}

$P(\nu_e \rightarrow \nu_e)$ is extremely simple as 
\begin{eqnarray}
P(\nu_e \rightarrow \nu_e)  &=&
1 - 
4 c_\phi^2 s_\phi^2 \sin^2 \frac{ (\lambda_{+} - \lambda_{-} ) L }{4E}, 
\label{P-ee}
\end{eqnarray}
where L is the baseline.
The reasons for the simplicity is discussed in depth in section~\ref{sec:compact-F}. $P(\nu_e \rightarrow \nu_\mu)$ and $P(\nu_e \rightarrow \nu_\tau)$ are given by 
\begin{eqnarray}
&& P(\nu_e \rightarrow \nu_\mu) = 4 c_\phi^2 s_\phi^2 s^2_{23} \sin^2 \frac{ (\lambda_{+} - \lambda_{-}) L}{ 4E } 
\nonumber\\
&+& 
4 \epsilon \dMsq c_{12} s_{12} c_{23} s_{23} c_{\phi} s_{\phi} \cos \delta
\nonumber \\
&\times&
\left[
\left\{
- c_{\phi} c_{\left( \phi - \theta_{13} \right)}  
\frac{ 1 }{ ( \lambda_{-} - \lambda_{0} )  }
+ s_{\phi} s_{\left( \phi - \theta_{13} \right)} 
\frac{ 1}{ ( \lambda_{+} - \lambda_{0} )  }
\right\}
\sin^2 \frac{ (\lambda_{+} - \lambda_{-}) L}{ 4E } 
\right.
\nonumber \\
&&\hspace*{-2mm} {} +
\left.
\left\{
c_{\phi} c_{\left( \phi - \theta_{13} \right)}  
\frac{ 1 }{ ( \lambda_{-} - \lambda_{0} )  }
+ s_{\phi} s_{\left( \phi - \theta_{13} \right)} 
\frac{ 1}{ ( \lambda_{+} - \lambda_{0} )  }
\right\}
\left\{
\sin^2 \frac{ (\lambda_{+} - \lambda_{0}) L}{ 4E } - \sin^2 \frac{ (\lambda_{-} - \lambda_{0}) L}{ 4E } 
\right\}
\right] 
\nonumber\\ 
&-& 8 \epsilon \dMsq c_{12} s_{12} c_{23} s_{23} c_{\phi} s_{\phi} \sin \delta 
\left\{
c_{\phi} c_{\left( \phi - \theta_{13} \right)}  
\frac{ 1 }{ ( \lambda_{-} - \lambda_{0} )  }
+ s_{\phi} s_{\left( \phi - \theta_{13} \right)} 
\frac{ 1}{ ( \lambda_{+} - \lambda_{0} )  }
\right\}
\nonumber\\ 
&\times&
\sin \frac{ (\lambda_{+} - \lambda_{-}) L }{ 4E } 
\sin \frac{ (\lambda_{-} - \lambda_{0}) L }{ 4E }
\sin \frac{ (\lambda_{0} - \lambda_{+}) L }{ 4E }.
\label{eq:P-emu}
\end{eqnarray}
\begin{eqnarray}
&& P(\nu_e \rightarrow \nu_\tau) =
4 c_\phi^2 s_\phi^2 c^2_{23} \sin^2 \frac{ (\lambda_{+} - \lambda_{-}) L}{ 4E } 
\nonumber\\
&-& 
4 \epsilon \dMsq c_{12} s_{12} c_{23} s_{23} c_{\phi} s_{\phi} \cos \delta
\nonumber \\
&\times&
\left[
\left\{
- c_{\phi} c_{\left( \phi - \theta_{13} \right)}  
\frac{ 1 }{ ( \lambda_{-} - \lambda_{0} )  }
+ s_{\phi} s_{\left( \phi - \theta_{13} \right)} 
\frac{ 1}{ ( \lambda_{+} - \lambda_{0} )  }
\right\}
\sin^2 \frac{ (\lambda_{+} - \lambda_{-}) L}{ 4E } 
\right.
\nonumber \\
&&\hspace*{-2mm} {} +
\left.
\left\{
c_{\phi} c_{\left( \phi - \theta_{13} \right)}  
\frac{ 1 }{ ( \lambda_{-} - \lambda_{0} )  }
+ s_{\phi} s_{\left( \phi - \theta_{13} \right)} 
\frac{ 1}{ ( \lambda_{+} - \lambda_{0} )  }
\right\}
\left\{
\sin^2 \frac{ (\lambda_{+} - \lambda_{0}) L}{ 4E } - \sin^2 \frac{ (\lambda_{-} - \lambda_{0}) L}{ 4E } 
\right\}
\right] 
\nonumber\\ 
&+& 8 \epsilon \dMsq c_{12} s_{12} c_{23} s_{23} c_{\phi} s_{\phi} \sin \delta 
\left\{
c_{\phi} c_{\left( \phi - \theta_{13} \right)}  
\frac{ 1 }{ ( \lambda_{-} - \lambda_{0} )  }
+ s_{\phi} s_{\left( \phi - \theta_{13} \right)} 
\frac{ 1}{ ( \lambda_{+} - \lambda_{0} )  }
\right\}
\nonumber\\ 
&\times&
\sin \frac{ (\lambda_{+} - \lambda_{-}) L }{ 4E } 
\sin \frac{ (\lambda_{-} - \lambda_{0}) L }{ 4E }
\sin \frac{ (\lambda_{0} - \lambda_{+}) L }{ 4E }.
\label{eq:P-etau}
\end{eqnarray}
It is almost trivial to verify unitarity in the $\nu_e -$row: 
$P(\nu_e \rightarrow \nu_e) + P(\nu_e \rightarrow \nu_\mu) + P(\nu_e \rightarrow \nu_\tau) = 1$.

\subsection{Oscillation probabilities in $\nu_\mu -$row}

$P(\nu_\mu \rightarrow \nu_e)$ is related to the T-conjugate channel probability $P(\nu_e \rightarrow \nu_\mu)$ as 
$P(\nu_\mu \rightarrow \nu_e: \delta) = P(\nu_e \rightarrow \nu_\mu: - \delta)$, 
whose latter can be obtained by replacing $\delta$ by $- \delta$ in (\ref{eq:P-emu}). Therefore, we only give the expressions of $P(\nu_\mu \rightarrow \nu_\mu)$ and $P(\nu_\mu \rightarrow \nu_\tau)$:  
\begin{eqnarray}
&& P(\nu_\mu \rightarrow \nu_\mu) 
\nonumber \\
&=& 
1 - 4 s^4_{23} c^2_{\phi} s^2_{\phi} 
\sin^2 \frac{ (\lambda_{+} - \lambda_{-} ) L }{ 4E } 
- 4 c^2_{23} s^2_{23} 
\left[
c^2_{\phi} \sin^2 \frac{ (\lambda_{+} - \lambda_{0} ) L }{ 4E } + 
s^2_{\phi} \sin^2 \frac{ (\lambda_{-} - \lambda_{0} ) L }{ 4E } 
\right]
\nonumber \\ 
&+& 
8 \epsilon \dMsq c_{12} s_{12} c_{23} s_{23} \cos \delta 
\nonumber \\ 
&\times& 
\left[ 
s^2_{23} c_{\phi} s_{\phi} 
\left\{ c_{\phi} c_{\left( \phi - \theta_{13} \right) }  
\frac{ 1 }{ ( \lambda_{-} - \lambda_{0} ) } 
- s_{\phi} s_{\left( \phi - \theta_{13} \right)}  
\frac{ 1 }{ ( \lambda_{+} - \lambda_{0} )  } 
\right\} 
\sin^2 \frac{ (\lambda_{+} - \lambda_{-} ) L }{ 4E } 
\right.
\nonumber \\ 
&&\hspace*{-6mm} {} - 
\left.
c_{\phi} 
\left\{ 
s^2_{23} s_{\phi} c_{\phi} c_{\left( \phi - \theta_{13} \right) }  
\frac{ 1 }{ ( \lambda_{-} - \lambda_{0} )  } 
+ \left( c^2_{23} - s^2_{23} c^2_{\phi}  \right) 
s_{\left( \phi - \theta_{13} \right)} 
\frac{ 1 }{ ( \lambda_{+} - \lambda_{0} )  } 
\right\} 
\sin^2 \frac{ (\lambda_{+} - \lambda_{0} ) L }{ 4E } 
\right.
\nonumber \\
&&\hspace*{-6mm} {} +
\left.
s_{\phi} 
\left\{ 
\left( c^2_{23} - s^2_{23} s^2_{\phi} \right) 
c_{\left( \phi - \theta_{13} \right)} 
\frac{ 1 }{ ( \lambda_{-} - \lambda_{0} ) } 
+ s^2_{23} c_{\phi} s_{\phi} s_{\left( \phi - \theta_{13} \right) }  
\frac{ 1 }{ ( \lambda_{+} - \lambda_{0} ) } 
\right\} 
\sin^2 \frac{ (\lambda_{-} - \lambda_{0} ) L }{ 4E } 
\right]. 
\nonumber \\
\label{eq:P-mumu}
\end{eqnarray}
\begin{eqnarray}
&& P(\nu_\mu \rightarrow \nu_\tau) 
\nonumber \\
&=&
4 c^2_{23} s^2_{23} 
\left[ 
- c_\phi^2 s_\phi^2 \sin^2 \frac{ (\lambda_{+} - \lambda_{-}) L}{ 4E } 
+ c_\phi^2 \sin^2 \frac{ (\lambda_{+} - \lambda_{0}) L}{ 4E }
+ s_\phi^2 \sin^2 \frac{ (\lambda_{-} - \lambda_{0}) L}{ 4E } 
\right] 
\nonumber \\
&+& 4 \epsilon \dMsq c_{12} s_{12} c_{23} s_{23} 
\left( c^2_{23} - s^2_{23} \right) \cos \delta 
\nonumber \\ 
&\times& 
\left[ 
c_{\phi} s_{\phi} 
\left\{ 
c_{\phi} c_{\left( \phi - \theta_{13} \right) }  
\frac{ 1 }{ ( \lambda_{-} - \lambda_{0} )  }
- s_{\phi} s_{\left( \phi - \theta_{13} \right)}  
\frac{ 1 }{ ( \lambda_{+} - \lambda_{0} )  } 
\right\}  
\sin^2 \frac{ (\lambda_{+} - \lambda_{-} ) L }{ 4E } 
\right.
\nonumber \\
&&\hspace*{-2mm} {} - 
\left.
\left\{ 
s_{\phi} c^2_{\phi} c_{\left( \phi - \theta_{13} \right)}  
\frac{ 1 }{ ( \lambda_{-} - \lambda_{0} )  } 
- c_{\phi} \left( 1 + c^2_{\phi} \right) s_{\left( \phi - \theta_{13} \right)}  
\frac{ 1 }{ ( \lambda_{+} - \lambda_{0} )  } 
\right\} 
\sin^2 \frac{ (\lambda_{+} - \lambda_{0} ) L}{ 4E }
\right.
\nonumber \\
&&\hspace*{-2mm} {} -
\left.
\left\{ s_{\phi} \left( 1 + s^2_{\phi} \right) c_{\left( \phi - \theta_{13} \right)}  
\frac{ 1 }{ ( \lambda_{-} - \lambda_{0} )  } 
- c_{\phi} s^2_{\phi} s_{\left( \phi - \theta_{13} \right)}  
\frac{ 1 }{ ( \lambda_{+} - \lambda_{0} )  }  \right\} 
\sin^2 \frac{ (\lambda_{-} - \lambda_{0} ) L }{ 4E }
\right]
\nonumber \\
&-&
8 \epsilon \dMsq c_{12} s_{12} c_{23} s_{23} c_{\phi} s_{\phi} \sin \delta 
\left\{
c_{\phi} c_{\left( \phi - \theta_{13} \right)}  
\frac{ 1 }{ ( \lambda_{-} - \lambda_{0} )  }
+ s_{\phi} s_{\left( \phi - \theta_{13} \right)} 
\frac{ 1}{ ( \lambda_{+} - \lambda_{0} )  }
\right\}
\nonumber\\ 
&\times& 
\sin \frac{ (\lambda_{+} - \lambda_{-}) L }{ 4E } 
\sin \frac{ (\lambda_{-} - \lambda_{0}) L }{ 4E }
\sin \frac{ (\lambda_{0} - \lambda_{+}) L }{ 4E }.
\label{eq:P-mutau}
\end{eqnarray}
With the above results the unitarity in $\nu_\mu -$row can also be verified: 
$P(\nu_\mu \rightarrow \nu_e) + P(\nu_\mu \rightarrow \nu_\mu) + P(\nu_\mu \rightarrow \nu_\tau) = 1$. 

\subsection{Oscillation probabilities in $\nu_\tau -$row}

$P(\nu_\tau \rightarrow \nu_e)$ and $P(\nu_\tau \rightarrow \nu_\mu)$ can be given by their T-conjugate channels: 
$P(\nu_\tau \rightarrow \nu_\alpha: \delta) = P(\nu_\alpha \rightarrow \nu_\tau: - \delta)$. Therefore, we only give the expressions of $P(\nu_\tau \rightarrow \nu_\tau)$ below.
\begin{eqnarray}
&& P(\nu_\tau \rightarrow \nu_\tau) 
\nonumber \\
&=&  
1 - 4 c^4_{23} c^2_{\phi} s^2_{\phi} 
\sin^2 \frac{ (\lambda_{+} - \lambda_{-} ) L }{ 4E } 
- 4 c^2_{23} s^2_{23} 
\left[
c^2_{\phi} \sin^2 \frac{ (\lambda_{+} - \lambda_{0}) L }{ 4E } + 
s^2_{\phi} \sin^2 \frac{ (\lambda_{-} - \lambda_{0}) L }{ 4E } 
\right]
\nonumber \\
&-& 
8 \epsilon \dMsq c_{12} s_{12} c_{23} s_{23} \cos \delta 
\nonumber \\ 
&\times& 
\left[ 
c^2_{23} c_{\phi} s_{\phi} 
\left\{ 
c_{\phi} c_{\left( \phi - \theta_{13} \right)}  
\frac{ 1 }{ ( \lambda_{-} - \lambda_{0} )  } 
- s_{\phi} s_{\left( \phi - \theta_{13} \right)}  
\frac{ 1 }{ ( \lambda_{+} - \lambda_{0} )  } 
\right\} 
\sin^2 \frac{ (\lambda_{+} - \lambda_{-} ) L }{ 4E } 
\right.
\nonumber \\ 
&&\hspace*{-6mm} {} -
\left.
c_{\phi} \left\{ 
c^2_{23} s_{\phi} c_{\phi} c_{\left( \phi - \theta_{13} \right) }  
\frac{ 1 }{ ( \lambda_{-} - \lambda_{0} )  } 
+ \left( s^2_{23} - c^2_{23} c^2_{\phi} \right) 
s_{\left( \phi - \theta_{13} \right)} 
\frac{ 1 }{ ( \lambda_{+} - \lambda_{0} )  } 
\right\} 
\sin^2 \frac{ (\lambda_{+} - \lambda_{0} ) L }{ 4E } 
\right.
\nonumber \\
&&\hspace*{-6mm} {} +
\left.
s_{\phi} 
\left\{ 
\left( s^2_{23} -  c^2_{23} s^2_{\phi} \right) 
c_{\left( \phi - \theta_{13} \right)} 
\frac{ 1 }{ ( \lambda_{-} - \lambda_{0} )  } 
+ c^2_{23} c_{\phi} s_{\phi} s_{\left( \phi - \theta_{13} \right)}  
\frac{ 1 }{ ( \lambda_{+} - \lambda_{0} )  }
\right\} 
\sin^2 \frac{ (\lambda_{-} - \lambda_{0} ) L }{ 4E } 
\right]. 
\nonumber \\
\label{eq:P-tautau}
\end{eqnarray}
The unitarity in $\nu_\tau -$row can also be verified: 
$P(\nu_\tau \rightarrow \nu_e) + P(\nu_\tau \rightarrow \nu_\mu) + P(\nu_\tau \rightarrow \nu_\tau) = 1$.

\section{Some useful Identities}
\label{sec:relations}

A straightforward derivation of the general expressions of the oscillation probabilities contains the following form of CP- or T-violation terms
\begin{eqnarray}
- 2 \sum_{j > i} 
{\mbox Im}[V_{\alpha i} V_{\beta i}^* V_{\alpha j}^* V_{\beta j}] 
\sin \frac{ (\lambda_{j} - \lambda_{i}) L}{ 2E }, 
\label{eq:CP-T-violation}
\end{eqnarray}
where $i = 0, -, +$ (or $1, 2, 3$). To cast this term to the one in (\ref{eq:P-ba-general}), one needs the following identity
\begin{eqnarray}
& & \sin \frac{ (\lambda_{+} - \lambda_{-}) L }{ 2E }
- \sin \frac{ (\lambda_{+} - \lambda_{0}) L }{ 2E }
+ \sin \frac{ (\lambda_{-} - \lambda_{0}) L }{ 2E }  \nonumber \\
& & \quad \quad \quad = - 4 
\sin \frac{ (\lambda_{+} - \lambda_{-}) L }{ 4E } 
\sin \frac{ (\lambda_{-} - \lambda_{0}) L }{ 4E }
\sin \frac{ (\lambda_{0} - \lambda_{+}) L }{ 4E }.
\label{eq:sss-identity}
\end{eqnarray}

Here, we list some more formulas which may be useful to understand the relationship between different expressions of the oscillation probabilities:
\begin{eqnarray} 
\cos 2 (\phi - \theta_{13}) = 
\frac{ \dMsq - \am ~ \cos 2\theta_{13}  }{( \lambda_{+} - \lambda_{-} )},  
\hspace{10mm}
\sin 2 (\phi - \theta_{13}) = 
\frac{ \am  ~\sin 2\theta_{13}  }{ (\lambda_{+} - \lambda_{-} )}. 
\label{eq:cos-sin-phi-13}
\end{eqnarray}
\begin{eqnarray}
c_{\phi} c_{\left( \phi - \theta_{13} \right)} &=& 
\frac{ c_{13}  }{ 2 ( \lambda_{+} - \lambda_{-} ) } 
\left[ ( \lambda_{+} - \lambda_{-} ) + (\dMsq  - \am) \right], 
\nonumber \\
s_{\phi} s_{\left( \phi - \theta_{13} \right)} &=& 
\frac{ c_{13}  }{ 2 ( \lambda_{+} - \lambda_{-} ) } 
\left[ ( \lambda_{+} - \lambda_{-} ) - ( \dMsq -  \am ) 
\right].
\label{eq:coeff-cc-ss}
\end{eqnarray}

\begin{eqnarray}
c_{\phi} s_{\left( \phi - \theta_{13} \right)} &=& 
\frac{ - s_{13}  }{ 2 ( \lambda_{+} - \lambda_{-} ) } 
\left[ (\lambda_{+} - \lambda_{-}) - ( \dMsq + \am ) \right], 
\nonumber \\
&= &\frac{ s_{13}  }{ ( \lambda_{+} - \lambda_{-} ) } 
\left[  \lambda_{-} -s^2_{12} \epsilon \dMsq \right], 
\nonumber \\ 
s_{\phi} c_{\left( \phi - \theta_{13} \right)} &=& 
\frac{ s_{13}  }{ 2 ( \lambda_{+} - \lambda_{-} ) } 
\left[ ( \lambda_{+} - \lambda_{-} ) + (\dMsq + \am) \right], 
\nonumber \\
&=& 
\frac{ s_{13}  }{ ( \lambda_{+} - \lambda_{-} ) } 
\left[  \lambda_{+} - s^2_{12} \epsilon \dMsq  \right]. 
\label{eq:coeff-cs-sc}
\end{eqnarray}

\begin{acknowledgments}
We thank Nordic Institute for Theoretical Physics (NORDITA) and the Kavli Institute for Theoretical Physics in UC Santa Barbara for their hospitalities, where part of this work was done. This research was supported in part by the National Science Foundation under Grant No. NSF PHY11-25915. 
 
H.M. thanks Universidade de S\~ao Paulo for the great opportunity of stay under ``Programa de Bolsas para Professors Visitantes Internacionais na USP''. He is grateful to Theory Group of Fermilab for supports and warm hospitalities during the several visits in 2012-2015 where this work was started and made corner turning progresses.

S.P.  acknowledges partial support from the  European Union FP7  ITN INVISIBLES (Marie Curie Actions, PITN- GA-2011- 289442). Fermilab is operated by the Fermi Research Alliance under contract no. DE-AC02-07CH11359 with the U.S. Department of Energy. 

\end{acknowledgments}


\begin{thebibliography}{99}

\bibitem{Wolfenstein:1977ue}
  L.~Wolfenstein,
 ``Neutrino Oscillations in Matter,''
  Phys.\ Rev.\ D {\bf 17} (1978) 2369.

\bibitem{Barger:1980tf}
  V.~D.~Barger, K.~Whisnant, S.~Pakvasa and R.~J.~N.~Phillips,
 ``Matter Effects on Three-Neutrino Oscillations,''
  Phys.\ Rev.\ D {\bf 22} (1980) 2718.

\bibitem{Zaglauer:1988gz}
  H.~W.~Zaglauer and K.~H.~Schwarzer,
 ``The Mixing Angles in Matter for Three Generations of Neutrinos and the MSW Mechanism,''
  Z.\ Phys.\ C {\bf 40} (1988) 273.

\bibitem{Kimura:2002wd}
  K.~Kimura, A.~Takamura and H.~Yokomakura,
 ``Exact formulas and simple CP dependence of neutrino oscillation probabilities in matter with constant density,''
  Phys.\ Rev.\ D {\bf 66} (2002) 073005
  [hep-ph/0205295].

\bibitem{Blennow:2013rca}
  M.~Blennow and A.~Y.~Smirnov,
 ``Neutrino propagation in matter,''
  Adv.\ High Energy Phys.\  {\bf 2013} (2013) 972485
  [arXiv:1306.2903 [hep-ph]].

\bibitem{Arafune:1997hd}
  J.~Arafune, M.~Koike and J.~Sato,
 ``CP violation and matter effect in long baseline neutrino oscillation experiments,''
  Phys.\ Rev.\ D {\bf 56} (1997) 3093
   [Erratum-ibid.\ D {\bf 60} (1999) 119905]
  [hep-ph/9703351].

\bibitem{Cervera:2000kp}
  A.~Cervera, A.~Donini, M.~B.~Gavela, J.~J.~Gomez Cadenas, P.~Hernandez, O.~Mena and S.~Rigolin,
 ``Golden measurements at a neutrino factory,''
  Nucl.\ Phys.\ B {\bf 579} (2000) 17
   [Erratum-ibid.\ B {\bf 593} (2001) 731]
  [hep-ph/0002108].

\bibitem{Arafune:1996bt}
  J.~Arafune and J.~Sato,
 ``CP and T violation test in neutrino oscillation,''
  Phys.\ Rev.\ D {\bf 55} (1997) 1653
  [hep-ph/9607437].

\bibitem{Freund:2001pn}
  M.~Freund,
 ``Analytic approximations for three neutrino oscillation parameters and probabilities in matter,''
  Phys.\ Rev.\ D {\bf 64} (2001) 053003
  [hep-ph/0103300].

\bibitem{Akhmedov:2004ny}
  E.~K.~Akhmedov, R.~Johansson, M.~Lindner, T.~Ohlsson and T.~Schwetz,
 ``Series expansions for three flavor neutrino oscillation probabilities in matter,''
  JHEP {\bf 0404} (2004) 078
  [hep-ph/0402175].

\bibitem{Mikheev:1986gs}
  S.~P.~Mikheev and A.~Y.~Smirnov,
 ``Resonance Amplification of Oscillations in Matter and Spectroscopy of Solar Neutrinos,''
  Sov.\ J.\ Nucl.\ Phys.\  {\bf 42} (1985) 913
   [Yad.\ Fiz.\  {\bf 42} (1985) 1441].

\bibitem{Nunokawa:2005nx}
  H.~Nunokawa, S.~J.~Parke and R.~Zukanovich Funchal,
 ``Another possible way to determine the neutrino mass hierarchy,''
  Phys.\ Rev.\ D {\bf 72} (2005) 013009
  [hep-ph/0503283].

\bibitem{An:2013zwz}
  F.~P.~An {\it et al.}  [Daya Bay Collaboration],
 ``Spectral measurement of electron antineutrino oscillation amplitude and frequency at Daya Bay,''
  Phys.\ Rev.\ Lett.\  {\bf 112} (2014) 061801
  [arXiv:1310.6732 [hep-ex]].

\bibitem{Jarlskog:1985ht}
  C.~Jarlskog,
 ``Commutator of the Quark Mass Matrices in the Standard Electroweak Model and a Measure of Maximal CP Violation,''
  Phys.\ Rev.\ Lett.\  {\bf 55} (1985) 1039.

\bibitem{Naumov:1991ju}
  V.~A.~Naumov,
 ``Three neutrino oscillations in matter, CP violation and topological phases,''
  Int.\ J.\ Mod.\ Phys.\ D {\bf 1} (1992) 379.

\bibitem{Harrison:1999df} 
  P.~F.~Harrison and W.~G.~Scott,
``CP and T violation in neutrino oscillations and invariance of Jarlskog's determinant to matter effects,''
  Phys.\ Lett.\ B {\bf 476}, 349 (2000)
  [hep-ph/9912435].

\bibitem{Asano:2011nj}
  K.~Asano and H.~Minakata,
 ``Large-Theta(13) Perturbation Theory of Neutrino Oscillation for Long-Baseline Experiments,''
  JHEP {\bf 1106} (2011) 022
  [arXiv:1103.4387 [hep-ph]].

\bibitem{Minakata:1998bf}
  H.~Minakata and H.~Nunokawa,
``CP violation versus matter effect in long baseline neutrino oscillation experiments,''
  Phys.\ Rev.\ D {\bf 57} (1998) 4403
  [hep-ph/9705208].

\bibitem{Agarwalla:2013tza}
  S.~K.~Agarwalla, Y.~Kao and T.~Takeuchi,
 ``Analytical approximation of the neutrino oscillation matter effects at large $\theta_{13}$,''
  JHEP {\bf 1404} (2014) 047
  [arXiv:1302.6773 [hep-ph]].

\bibitem{Kuo:1989qe}
  T.~K.~Kuo and J.~T.~Pantaleone,
 ``Neutrino Oscillations in Matter,''
  Rev.\ Mod.\ Phys.\  {\bf 61} (1989) 937.

\end{thebibliography}
\end{document}